\documentclass[screen,acmsmall]{acmart}

\AtBeginDocument{%
  \providecommand\BibTeX{{%
    \normalfont B\kern-0.5em{\scshape i\kern-0.25em b}\kern-0.8em\TeX}}}

\setcopyright{acmlicensed}
\acmJournal{PACMHCI}
\acmYear{2025} \acmVolume{9} \acmNumber{7} \acmArticle{CSCW506} \acmMonth{11}\acmDOI{10.1145/3757687}

\newcommand{\revision}[1]{{\textcolor{black}{#1}}}

\usepackage{subfig}
\usepackage{tabulary}
\usepackage{pdfpages}
\usepackage{ragged2e}

\begin{document}




\title[]{Understanding Data Usage when Making High-Stakes Frontline Decisions in Homelessness Services}


\author{Teale W. Masrani}
\email{teale.masrani2@ucalgary.ca}
\orcid{0000-0001-7749-1379}
\affiliation{%
  \institution{Computer Science, University of Calgary}
  \country{Canada}
}

\author{Geoffrey Messier}
\email{gmessier@ucalgary.ca}
\orcid{0000-0002-9825-3238}
\affiliation{%
  \institution{Electrical and Software Engineering, University of Calgary}
   \country{Canada}
}

\author{Amy Voida}
\email{amy.voida@colorado.edu}
\orcid{0000-0002-2490-2063}
\affiliation{%
  \institution{Information Science, University of Colorado Boulder}
   \country{USA}
}

\author{Gina Dimitropoulos}
\email{gdimit@ucalgary.ca}
\orcid{0000-0001-9487-0290}
\affiliation{%
  \institution{Social Work, University of Calgary}
  \country{Canada}
}

\author{Helen Ai He}
\email{helen.he1@ucalgary.ca}
\orcid{0000-0001-8681-2381}
\affiliation{%
  \institution{Computer Science, University of Calgary}
  \country{Canada}
}

\renewcommand{\shortauthors}{Teale W. Masrani, Geoffrey Messier, Amy Voida, Gina Dimitropoulos, \& Helen Ai He}


\begin{abstract}
Frontline staff of emergency shelters face challenges such as vicarious trauma, compassion fatigue, and burnout. The technology they use is often not designed for their unique needs, and can feel burdensome on top of their already cognitively and emotionally taxing work. While existing literature focuses on data-driven technologies that automate or streamline frontline decision-making about vulnerable individuals, we discuss scenarios in which staff may resist such automation. We then suggest how data-driven technologies can better align with their human-centred decision-making processes. This paper presents findings from a qualitative fieldwork study conducted from 2022 to 2024 at a large emergency shelter in Canada. The goal of this fieldwork was to co-design, develop, and deploy an interactive data-navigation interface that supports frontline staff when making collaborative, high-stakes decisions about individuals experiencing homelessness. By reflecting on this fieldwork, we contribute insight into the role that administrative shelter data play during decision-making, and unpack staff members' apparent reluctance to outsource decisions about vulnerable individuals to data systems. Our findings suggest a ``data-outsourcing continuum,'' which we discuss in terms of how designers may create technologies to support compassionate, data-driven decision-making in nonprofit domains.

\end{abstract}



\maketitle

\section{Introduction}
Nonprofit organizations collect large volumes of administrative data about their clients. These datasets can be used to evaluate programs, analyse client populations, and request funding, among other purposes \cite{mayer_exploring_2023}. Existing literature on nonprofit data tends to focus on how entities outside the organization can benefit from such data (e.g., \cite{bopp_disempowered_2017, culhane_potential_2016, mayer_exploring_2023}). In contrast, this study presents findings on \revision{how data is used inside the nonprofit organization, in client-facing, frontline roles, when staff make high-stakes decisions about vulnerable populations.}



This research centres on the cooperative design (henceforth ``co-design'') and deployment of a human-centred data-navigation interface for staff at the largest emergency shelter in Calgary, Canada: The Calgary Drop-In Centre. \revision{This shelter serves nearly 7000 unique clients per year, all of whom are actively experiencing homelessness -- a circumstance affecting approximately 40,000 people across Canada each night.} Frontline staff used this new interface to sift through administrative data while making high-impact, time-sensitive, and collaborative decisions about shelter clients. Specifically, \revision{the Drop-In Centre maintains a committee of staff dedicated to deciding how to move forward with clients who have been temporarily ``barred'' from entering the shelter due to inappropriate or dangerous misconduct. This committee used the new, custom-built data-navigation interface during their weekly discussions.}

During this \revision{fieldwork}, we iteratively designed and deployed a prototype of the interface while conducting semi-structured interviews, interactive co-design workshops, and ethnographic observations. In the remainder of this paper, we use the terminology ``homeless'' or ``person experiencing homelessness'' to refer to \revision{those living without safe or stable housing, and who are unable to obtain it due to systemic or economic challenges}. We use the term ``vulnerability'' when describing the degree to which a person is in need of community services. Although some recent work has debated such terms \cite{garrett_vulnerabilized_2024, tsai_terms_2023}, we chose to use these terms as they match the language used by the frontline staff at this shelter. 

Our previously published work highlighted preliminary findings from this co-design study. It characterized the importance of data-driven technologies that help develop an understanding of \textit{the human behind the data}, in spite of the inherent pitfalls of administrative data such as bias that interferes with understanding client circumstances \cite{masrani_human_2023}. 
This paper extends our early work by using reflexive thematic analysis \cite{braun_reflecting_2019, braun_toward_2023, byrne_worked_2022}, within a design ethnography framework \cite{baskerville_design_2015}, to answer the following research questions: \textit{How do staff perceive the role of administrative data when deciding how to serve those experiencing homelessness? How can designers incorporate this staff--data relationship when creating technologies to support staff who make high-stakes frontline decisions?} 

In answering these questions, we make the following contributions to the literature. First, we describe our \revision{18-month fieldwork study, explaining the} process of co-designing a collaborative decision-making interface with frontline staff, and we analyse how staff experienced using it over one year. Second, we investigate the role of administrative shelter data in frontline decision-making about vulnerable clients, unpacking the benefits and challenges of relying on data in high-stakes scenarios where decision \revision{outcomes} can greatly affect clients' lives. Third, we articulate a ``data-outsourcing continuum'' \revision{which underlies staff members' relationships with data-driven decision-making, and their willingness to ground their decisions in data systems.} We additionally foreground key questions for designers to consider when creating computational decision-aid tools for \revision{care work in} nonprofit environments.

\section{Background and Related Work}
\label{sec:RelatedWork}
This section first provides a background on homelessness, the role that emergency shelters play in combatting homelessness, and the challenges faced by frontline staff at emergency shelters. We then present previous work on technology and data in community-based nonprofit organizations. Finally, we present literature on technology and data specifically within emergency shelters.

\subsection{Homelessness and the Role of Emergency Shelters}

On a single night, over 40,000 individuals experience homelessness in Canada \cite{government_of_canada_everyone_2024}, and over 600,000 do in the United States \cite{soucy_state_2024}.
The United Nations defines homelessness as ``not having stable, safe and adequate housing, nor the means and ability of obtaining it'' \cite{united_nations_homelessness_2024}. A person who is experiencing homelessness has likely faced multiple acute or chronic injustices. Intimate partner violence, chronic mental or physical health ailments, substance use disorders, and continuous systemic discrimination are among several factors that can lead to homelessness, especially when the societal context does not provide adequate resources to overcome these challenges \cite{calgary_homeless_foundation_causes_2023}. Notably, homelessness can also \textit{cause} these challenges. Hence, homelessness spans a wide spectrum of overlapping issues, making it a complex and cyclical phenomenon that goes beyond the absence of housing. 


This circumstance can rapidly harm a person's psychological and physical health. Psychologically, they may feel intense shame and anxiety about their life circumstances as they face a sudden and ongoing loss of basic daily safety and security. Compared to those with housing, homelessness correlates with higher rates of mortality, infectious diseases, sexual health issues, long-term mental and physical health conditions, and substance use disorders \cite{zaretzky_what_2017, jaworsky_residential_2016, fazel_health_2014, sleet_homelessness_2021, schwartz_sexual_2023, hossain_prevalence_2020, zaretzky_what_2017}. These challenges, coupled with systemic obstacles, inhibit an individual's ability to meet their basic needs and escape homelessness. Further, public health infrastructure also suffers due to increased healthcare resource utilization \cite{zaretzky_what_2017, jaworsky_residential_2016}. For example, having no home to recover in severely limits one's ability to heal after a health emergency, leading to repeated hospital visits \cite{sleet_homelessness_2021}. These multifaceted impacts \revision{on individuals and society} call for immediate interventions.

\subsection{Challenges Faced by Frontline Staff}
Emergency shelters offer short-term relief with a place to escape the outdoors and have a meal. Many shelters also offer services to help clients meet their long-term goals, such as physical and mental health care, professional case management, and programs to connect clients with affordable, long-term housing. This research took place in an emergency shelter that provides such services. 

When an individual enters a shelter, they are often in an exceptionally compromised state. Every client exhibits unique needs, and dedicated staff must build relationships with each person to develop a personalized support plan. Unfortunately, staff encounter many challenges in this work environment, often suffering from mental health issues themselves. For example, they grapple with vicarious trauma and compassion fatigue due to continuous exposure to traumatic events and clients' extreme distress \cite{levesque_understanding_2021, olivet_staffing_2010, waegemakers_schiff_ptsd_2019}. They can also face verbal or physical abuse from clients who act erratically \cite{levesque_understanding_2021, tanjuaquio_san_2018}. These challenges significantly increase cognitive and emotional loads for staff, contributing to ongoing stress, disillusionment, burnout, and high turnover \cite{levesque_understanding_2021, yamada_role_2022, waegemakers_schiff_ptsd_2019, lenzi_factors_2021, olivet_staffing_2010}. 

\revision{Staff typically come from a care work or community-focused career background, such as social work or nursing \cite{levesque_understanding_2021}. Within the shelter, they can fulfill roles in case management, security, healthcare, or several other areas.} Among their many responsibilities, staff must also effectively interact with digital tools. Technical skills are necessary for homelessness service jobs \cite{levesque_understanding_2021}, although nonprofit staff often have limited technological literacy \cite{mayer_exploring_2023}. As staff already contend with high emotional and cognitive loads, using burdensome technology that does not align with their values and workflows further contributes to burnout. Consequently, designing technology that supports frontline staff is critical for the ultimate goal of helping shelter clients.

\subsection{Technology and Data Use in Community-Based Nonprofits}
The nonprofit environment is uniquely challenging for human-centred computing, as nonprofit organizations' goals evolve rapidly at the whim of public and private sectors \cite{voida_shapeshifters_2011, boris_roles_2021, van_til_nonprofit_1994}. For example, when governments change which community services they provide, nonprofits' missions correspondingly change as they ``shapeshift'' to fill gaps and serve populations that are underprovided for by the public sector \cite{voida_shapeshifters_2011, boris_roles_2021, van_til_nonprofit_1994}. Additionally, resources and funding are constantly in flux \cite{voida_shapeshifters_2011, van_til_nonprofit_1994}, requiring nonprofits to remain highly flexible. Therefore, off-the-shelf technological solutions rarely meet their needs \cite{voida_shapeshifters_2011}. 
Along these lines, ample research has looked at designing effective technological tools for nonprofits (e.g. \cite{hackler_strategic_2007, mcnutt_technology_2018, bobsin_value_2019, merkel_managing_2007, voida_homebrew_2011, voida_shapeshifters_2011, seguin_co-designing_2022}), with much of it showing that participatory methods are important for staff to claim agency over their technology and ensure it reflects their core values and changing organizational needs \cite{seguin_co-designing_2022, merkel_managing_2007, voida_shapeshifters_2011}. This empowerment is instrumental for the nonprofit sector as its funding and resource constraints do not allow frequent investment in new technological tools. \revision{Further, due to the demanding nature of their work, staff do not often have the time or energy to incorporate new technological practices into their workflows.} 

The present work focuses specifically on how staff use technology to engage with shelter data. 
Although participatory methods, as noted, are crucial for staff empowerment, the literature lacks real-world accounts of collaborations with nonprofits where the aim is to support staff's engagement with data -- a gap this study fills. 
Further, despite the above literature urging the design of tools that empower staff, numerous ineffective approaches to handling data instead \textit{disempower} them \cite{voida_homebrew_2011, bopp_disempowered_2017}. First, nonprofit data are not typically organized in sustainable ways, causing staff to rely on patchwork solutions to store large amounts of disparate data \cite{voida_homebrew_2011}. Second, nonprofits struggle to standardize their data practices due to unreliable collection methods and minimal in-house technical expertise \cite{mayer_exploring_2023}. Third, nonprofit staff experience a cycle of disengagement and disempowerment as they attempt congruence with the organization's values, but find that external pressures from funders demand disruptive data collection practices \cite{bopp_disempowered_2017}. Lastly, a tension lies in nonprofit values that deprioritize data practices in favour of human-centred work 
\cite{mayer_exploring_2023, voida_shapeshifters_2011, herbert_towards_2015, slota_caring_2023}.  

Considerable existing findings focus on the needs of entities external to the nonprofit organization, such as funding bodies, thereby drawing attention to how data practices largely appear to hinder nonprofit goals \cite{mayer_exploring_2023}. As Mayer and Fischer (2023) suggest, deeper theorizing about data usage for \textit{intra}-organizational processes is needed \cite{mayer_exploring_2023}. Our research pursues these aims by unpacking the relationship that frontline staff have with data in a nonprofit environment; moreover, we generate a paradigm for understanding the degree to which staff members use data for critical decision-making about vulnerable clients.


\subsection{Technology and Data Use in Homelessness and Emergency Shelters}
\label{sec:RWHomelessSector}
We now shift from the general context of nonprofits to the specific context of homelessness \revision{services}. Most research at the intersection of technology and homelessness focuses on shelter clients as the users of technology (e.g. \cite{woelfer_homeless_2011, woelfer_improving_2011, mohan_food-availability_2019, le_dantec_designs_2008, koepfler_we_2012, chandra_critical_2021, heaslip_use_2021, calvo_information_2019}), investigating  how technology can support their health, safety, and well-being.
Some work has also explored technology use by both clients and staff within a shelter \cite{moser_text_2009, hendry_how_2011, le_dantec_tale_2010, le_dantec_publics_2011}, to navigate services, facilitate communication, and build relationships. The present work, however, considers frontline staff as the sole user group.  

Prior work has investigated how to leverage administrative data to conduct broad, population-level analyses of those experiencing homelessness (e.g., \cite{metraux_analyzing_1999, culhane_potential_2016, singham_discrete-event_2023, messier_best_2022, brush_data_2016, thomas_principles_2020, wachter_predicting_2019}) or to train machine learning (ML) models to identify predictors of homelessness (e.g. \cite{rahman_bayesian_2023, hong_applications_2018,tabar_identifying_2020, messier_predicting_2022}). However, shelter data can also be beneficial for day-to-day decision-making and program delivery. 
Many researchers overlook the\revision{se} ground-level applications of data within shelters, although some do investigate ML, predictive modelling, and recommendation systems to aid in the triage of resources to clients \cite{showkat_who_2023, kithulgoda_predictive_2022, chelmis_challenges_2021, chelmis_smart_2021, chan_evidence_2017, kuo_understanding_2023}. These triage tools allocate resources by ranking clients' levels of vulnerability. However, using such automated computational tools can present challenges for frontline staff. For example, according to Showkat et al.'s (2023) systematic review, the most common limitation to using algorithmic data tools for homelessness service provision is data quality issues, such as sample bias in the datasets and inaccurate self-reported data \cite{showkat_who_2023}. 

Two studies are of note with respect to how data tools can contribute to making decisions about vulnerable clients. Kuo et al. \cite{kuo_understanding_2023} studied staff and clients' perceptions of artificial intelligence (AI) tools that are trained on administrative shelter data, finding that staff have an acute understanding of how such data can and cannot be used to assess clients' vulnerability and make frontline decisions.
Additionally, Karusala et al. \cite{karusala_street-level_2019} explored the ``street-level realities'' of non-computational data tools such as the VI-SPDAT (Vulnerability Index--Service Prioritization Decision Assistance Tool) \cite{orgcode_consulting_inc_and_community_solutions_vulnerability_2015}. The VI-SPDAT was a simple and widely-used assessment tool in the homelessness sector for several years that relied on questionnaire data self-reported from clients to numerically rank their vulnerability. 
Domain experts and researchers pushed back against the VI-SPDAT, arguing that it oversimplified clients' circumstances  \cite{cronley_invisible_2022, brown_reliability_2018}. In December 2020, it began to be phased out \cite{orgcode_message_2020}. Karusala et al. found that these tools did not allow for staff to feel as though they could use them with integrity, in ways that honoured the needs of clients \cite{karusala_street-level_2019}. This was \revision{especially detrimental since} clients already lack agency over the decisions made on their behalf \revision{within emergency shelters.} 

These frontline data tools privilege quantitative data, the ``objectivity'' of which staff sometimes appreciate. Still, staff value narrative qualitative data, such as case notes, to better understand clients' needs \cite{karusala_street-level_2019, kuo_understanding_2023}. Bopp et al. \cite{bopp_showing_2023} foreground this sentiment, explaining that rich qualitative data about individual client experiences are imperative when measuring homelessness at the population level. These studies emphasize the importance of preserving the ``human touch'' \cite{karusala_street-level_2019} when attempting to understand clients through their data. The present research extends this body of work by identifying benefits and challenges staff encounter when \revision{working with} administrative data, which influence how willing they are to \revision{let such data drive their decision-making processes.}

\section{Research Context: The Calgary Drop-In Centre}
\label{sec:CalgaryDropInCentre}
This fieldwork took place at The Calgary Drop-In Centre: a large emergency shelter in Calgary, with 6839 unique individuals who accessed it from April 1, 2022, to March 31, 2023 \cite{calgary_drop-in_and_rehabilitation_centre_calgary_2023}. The Drop-In Centre helped 553 people find stable housing during this same period. \revision{The shelter is far busier during winter months, and} an average of 505 people sleep there nightly \cite{calgary_drop-in_and_rehabilitation_centre_calgary_2023}.
The Drop-In Centre has grown since 1961 from a basic overnight shelter into a hub for thousands of people to not only access a bed, but also \revision{use} specialized services \cite{calgary_drop-in_and_rehabilitation_centre_celebrating_2021}.

The building has six floors to accommodate check-in areas, a dining centre where volunteers serve three meals per day, a ``housing hub'' where clients meet with housing specialists, several bunk bed rooms, a healthcare centre with licensed medical professionals, a staffed detox room, and office spaces for employees. Throughout this paper, we refer to anyone seeking assistance from staff, whether they stay overnight or not, as a ``client.'' Clients range from 18 to over 65 years old. They occupy the common areas throughout the building; some are alert, social, and in good health, while many others are experiencing physical and mental health crises.
Client-facing frontline staff ensure that all individuals are safe and attended to, regardless of the state they are in.



The Drop-In Centre staff \revision{is vast, with over 300 full-time employees} comprising numerous teams, such as healthcare workers, leadership staff, impact analysts, the IT team, the shelter team, and the housing team. The latter two play the most central roles. The \textit{shelter team} is responsible for everyday operations, consisting of \textit{security staff} and \textit{emergency shelter workers}. While security staff protect the safety of everyone on the property, emergency shelter workers act as ``floaters'' throughout the building, to check people in, answer questions, and direct clients throughout the building, among other tasks. 
The \textit{housing team} helps clients find stable housing, and consists of two roles. \textit{Diversion workers} engage heavily with first-time clients during their first 21 days in shelter to ``divert'' them from the shelter and into housing as quickly as possible. Meanwhile, \textit{housing support workers} work continuously with complex clients who face chronic barriers that make it particularly difficult to escape homelessness. 

\subsection{The Bar Review Committee}
To investigate our research questions, we collaborated with one of the most high-stakes decision-making contexts at the shelter: the work of the Bar Review Committee (BRC). The BRC makes decisions about \textit{barring}. A staff member may \textit{bar} a client from using the shelter if they break building guidelines, for example by endangering the safety of others. Barred clients are disallowed from entering the shelter for a specified period of time -- a troubling experience for all involved. Bar types are organized into a hierarchy, ranging from categories 1 to 4. 
Category 1 bars may last only 24 hours, and would be placed after a minor physical altercation, using drugs in the building, verbal abuse, or other minor safety threats. Category 4 bars can last multiple months, and would be placed after extreme threats to safety, such as harm with a weapon or predatory behaviour. Categories 2 and 3 lie in between these two poles. Staff can also place a ``Long-Term Safety'' bar, which may last indefinitely, if a client has become a significant risk to others by engaging in extremely harmful behaviours such as sexual assault.

Although bars may seem critical for upholding shelter safety, they are highly distressing as life-saving warmth and services are withheld from the barred client, which is especially problematic during cold winter months with dangerously low temperatures. Thus, barring is a controversial topic among staff. Some believe that no client should ever be barred, some believe barring policies should be strictly followed, and others believe the barring system should improve but are unsure how to achieve this. Most staff also stress that barring is not punitive, but purely motivated by safety. Decisions about barring have high stakes and are difficult to make, especially if a client engages in illegal behaviours, causing staff to grapple with whether or not to involve police services as well. Therefore, in an effort to uphold fairness and integrity in the barring process, the Drop-In Centre maintains an active Bar Review Committee (BRC).

BRC members are the main participants of this research. They encompass a cross-section of shelter coordinators, security team coordinators, healthcare specialists, and housing team members. The BRC holds weekly meetings, during which they review Category 3 and 4 bars. Only the BRC staff are present at these meetings. For each client who received one of these high-category bars in the past week, the BRC discusses the barring circumstances, assesses the legitimacy and morality of the bar placement, and decides how to move forward with the client -- a difficult and emotionally charged task. They may decide to change the category or duration of the bar, replace the bar with a ``condition of entry'' such as speaking with a supervisor before reentering, lift the bar altogether, or keep the bar unchanged if they agree that it is fair. Meetings last two hours or more, depending on the complexity of each case. For every barred client they discuss, BRC members review a wide array of administrative data to comprehensively understand the situation. 

\subsection{Data and Technology at the Drop-In Centre}

As mentioned, over the course of a year, the staff at the Drop-In Centre serve close to 7000 unique individuals. 
As some clients may continue to interact with the shelter for more than a decade, they could have hundreds of administrative data points collected about them. All these data are organized into a relational database containing over 50 interrelated data tables, such as the \textit{building check-ins} table or the \textit{bars} table. The database is housed using Microsoft's cloud-based data storage software, \textit{Dataverse} \cite{microsoft_innovate_2023}. Staff read and write to the database via a graphical database management system interface referred to as ``Guestbook,'' which was built by the in-house IT team using Microsoft Dynamics 365 \cite{microsoft_what_2023}. 

The Drop-In Centre offered a fruitful \revision{case study} for this research endeavour given its large client base and how closely staff scrutinize administrative data to holistically understand client circumstances. Rich staff--client relationships can be difficult to develop given such a large number of clients, and staff must therefore defer to the data to learn about clients' stories.
The existing in-house IT infrastructure \revision{and database management system, Guestbook, also made for a unique shelter environment} to uncover insights about staff--data interactions \revision{in the homelessness context.}

Guestbook allows staff to manage client data in their day-to-day work, but it was not designed specifically for any one job function within the Drop-In Centre. When using Guestbook during BRC meetings, staff encounter numerous data-interaction obstacles. For instance, data feel scattered as they are displayed across 23 different pages, with no visual hierarchy to organize the data as it is displayed in several aesthetically-identical flattened data tables.
Therefore, the Drop-In Centre invited us to \revision{build upon the existing IT infrastructure to} create a \revision{new, read-only} tool that extends Guestbook's functionality, for \revision{more effective} data navigation during BRC discussions.



\section{Methods}
\label{sec:Methods}

With frontline staff at the Calgary Drop-In Centre, we co-designed a new data-navigation interface to facilitate weekly BRC meetings. Hence, we adopted a \textit{design ethnography} approach throughout the 18-month duration of this work.
While traditional ethnography positions the researcher as a ``student'' of a foreign culture or domain \cite{baskerville_design_2015, randall_ethnography_2014}, researchers in HCI disciplines not only learn from those in the field, but also become advisors and teachers, sharing technical expertise as they collaborate with domain experts \cite{baskerville_design_2015}. In 1997, Salvador and Mateas introduced the ``design ethnography'' methodology \cite{salvador_introduction_1997}, and in 2015, Baskerville and Myers discussed design ethnography in the context of designing information systems \cite{baskerville_design_2015}. 

\begin{figure} [h!]
    \centering
    \includegraphics[width=1\linewidth]{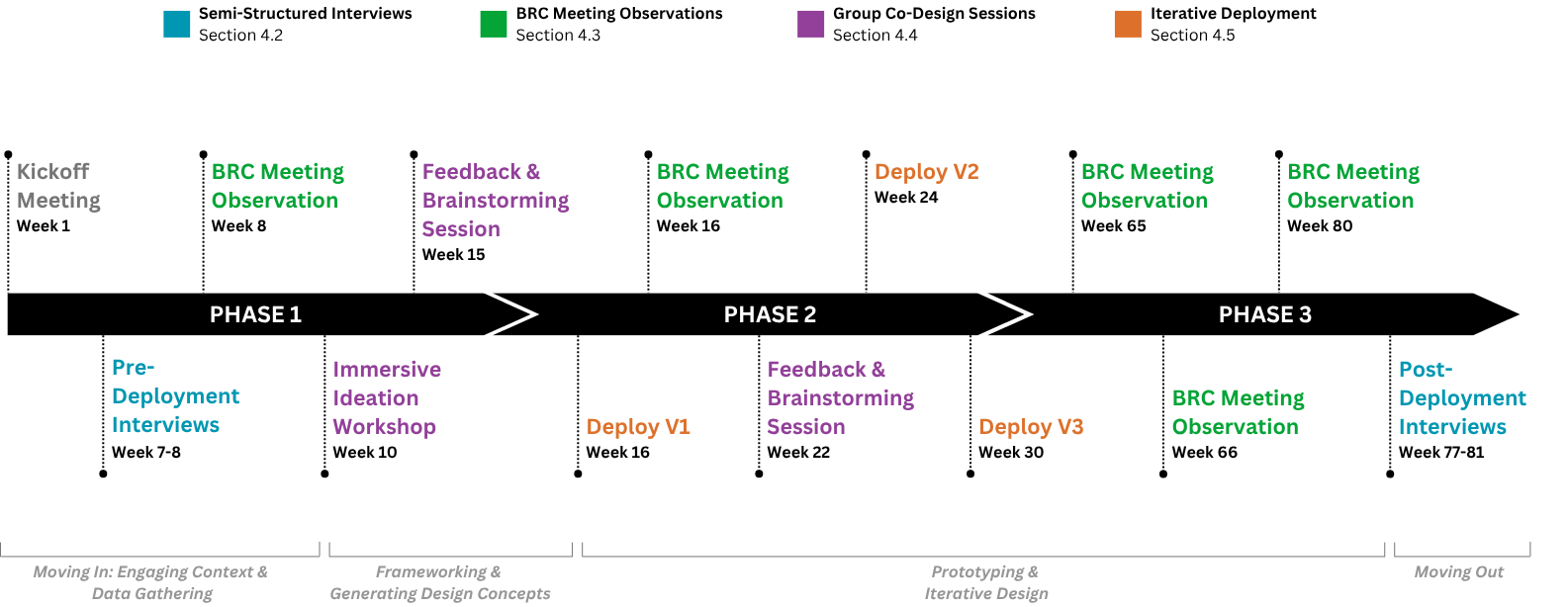}
    \caption{Timeline of all research events in the study with corresponding paper section numbers and week numbers, and the superimposed stages of the \textit{design ethnography} framework \cite{baskerville_design_2015}.}
    \label{fig:Timeline}
    \Description{A horizontal timeline diagram with three phases and several research events. Phase one includes the following events: kickoff meeting, pre-deployment interviews, BRC meeting observation, immersive ideation workshop, and feedback and brainstorming session. Phase two includes the following events: deploy V1, BRC meeting observation, feedback and brainstorming session, deploy V2, and deploy V3. Phase three includes the following events: three BRC meeting observations and post-deployment interviews. Below the timeline are the stages of the design ethnography framework shown parallel to the timeline: (1) moving in: engaging context and data gathering, (2) frameworking and generating design concepts, (3) prototyping and iterative design, and (4) moving out.}
\end{figure}

According to the design ethnography framework \cite{baskerville_design_2015}, there should be two parallel research threads: (i) the \textit{design} thread, resulting in practical IT interventions and artifacts, and (ii) the \textit{design ethnography} thread, resulting in a descriptive \textit{ethnography} that contributes new knowledge to the field. The present work adopts this methodology to create a data-navigation interface for the BRC, while also ethnographically studying the emergency shelter design context. This research included semi-structured interviews, ethnographic BRC meeting observations, group co-design sessions, and iterative deployments of the new platform. All events were approved by the University of Calgary Conjoint Faculties Research Ethics Board (REB21-1121), and are shown in Figure \ref{fig:Timeline} alongside the design ethnography framework. See \cite{baskerville_design_2015} for the original design ethnography framework diagram. 

\begin{table} [ht]
    \centering
    \resizebox{\textwidth}{!}{
    \begin{tabular}{p{1cm} p{2.25cm} p{2.25cm} p{2.25cm} p{2.25cm} p{2.25cm} p{2.25cm}} 
         \RaggedRight{ID} & \RaggedRight{Pre-\newline Deployment \newline Interviews}& \RaggedRight{Immersive \newline Ideation \newline Workshop}& \RaggedRight{Feedback \& \newline Brainstorming \newline Session (V0)}& \RaggedRight{Feedback \&  \newline Brainstorming \newline Session (V2)}& \RaggedRight{Post-\newline Deployment \newline Interviews}& \RaggedRight{BRC Meetings Attended \newline(of 5 observed)}\\ \hline 
         P01& X& X& X& X& & 2/5\\  \hline
         P02& X& X& X& X& & 1/5\\  \hline
         P03& X& & & & & 1/5\\ \hline
         P04& X& X& X& & & 1/5\\ \hline
         P05& X& X& X& X& X& 5/5\\  \hline
         P06& & & & X& X& 5/5\\  \hline
         P07& & & X& X& &\\  \hline
         P08& & & & X& &\\ \hline
         P09& & & & & X& 3/5\\  \hline
         P10& & & & & X& 3/5\\  \hline
         P11& & & & & X& 2/5\\  \hline
         P12& & & & & X& 2/5\\  \hline
         P13& & & & & X& 3/5\\  \hline
         P14\_IT& & X& X& X& &\\ \hline
         P15\_IT& & X& X& X & &\\ \hline
         P16\_IT& & X& & & &\\ \hline
    \end{tabular}}
    \caption{Participants Table, including participant IDs and their involvement in each research event.}
    \label{tab:participants}
\end{table}

\subsection{Participants}
Table \ref{tab:participants} lists participants via their anonymous IDs and which events they participated in. \revision{In total, 16 staff members at the Drop-In Centre were involved in this collaboration.} The cohort consisted of six women and ten men, \revision{all under the age of 40,} with four shelter leadership members, two emergency shelter workers, two security team members, two housing team members, two impact analysts, one healthcare worker, and three members of the IT team. \revision{Given that the BRC is a small group of staff, we have intentionally excluded further description of their demographic background to protect participant anonymity.}
%
The daily work in homeless services can be demanding and unpredictable \cite{mcdonald_mental_2021}. Therefore, it was expected that not all participants would participate in every research event, as depicted in the participant table. 
Furthermore, BRC membership evolved over time, causing some staff to become less involved and others more involved over the course of this research.
\revision{Specifically, P1, P2, P3, and P4 were only involved in the BRC when the research began, taking part in shaping the initial interface design. Later, P9, P10, P11, P12, and P13 became more active in the BRC, contributing to insights generated in Phase 3. P7 and P8 were not involved in the BRC and only chose to participate briefly during co-design sessions, whereas P5 and P6 were active during all BRC meetings.}



\subsection{Semi-Structured Interviews}
\label{sec:interviews}
The \textit{Moving In} stage of the design ethnography framework \cite{baskerville_design_2015} began during the initial kickoff meeting in June, 2022. Then, as depicted in Figure \ref{fig:Timeline}, the pre-deployment interviews corresponded with the \textit{Engaging Context} and \textit{Data Gathering} design ethnography stages. In these stages, researchers gather data that inform practical aspects of designing the intended artifacts \cite{baskerville_design_2015}. Accordingly, the goal of the pre-deployment interviews was to understand participants' existing data needs in regards to barring, and converge on a vision for the new interface. Each one-on-one interview was 45 to 60 minutes long. 
\revision{Interviewees shared experiences of their typical workflows at the Drop-In Centre and their thoughts on data in the barring and BRC process. These interviews surfaced participants' values and needs, illuminating initial goals for the new interface.}

As seen in Figure \ref{fig:Timeline}, this research concluded 16 months later with post-deployment interviews, each also one-on-one and 45 to 60 minutes long.
\revision{For this round of interviews, participants shared their evaluations of the new interface, recalling their experiences over the past year, and discussed their thoughts on client data usage at the Drop-In Centre.}
Responses informed high-level takeaways regarding the benefits and challenges of incorporating data into frontline work, and when data-driven computational tools may or may not be most helpful. 
\revision{See Appendix \ref{appendixA} for pre-deployment and post-deployment interview guides. While each set of interview questions provided a framework of topics to discuss, interviews were guided by what was deemed meaningful for participants, in accordance with a semi-structured interview protocol \cite{byrne_worked_2022}.}

\subsection{Bar Review Committee (BRC) Meeting Observations}
\label{sec:BRCObservations}
Soon after finishing pre-deployment interviews, two researchers sat in on a BRC meeting. While \revision{the} pre-deployment interviews revealed participants' perspectives of how data is used when reviewing bars, this BRC meeting observation deepened our understanding with a first-hand account of the use case. This observation filled gaps in interview data by providing concrete examples of BRC discussions and data-usage. During this meeting, staff used their typical Guestbook platform to sift through data tables and collaboratively decide how to move forward with each client. The two observing researchers did not \revision{speak or interject during the} meeting. \revision{Anonymous} field notes were taken throughout the observation.

A second BRC meeting observation took place after the first deployment of the new interface. 
Consistent with the design ethnography methodology \cite{baskerville_design_2015}, \revision{the researchers} acted not only as ethnographers but also as technical advisors to assist with any system integration issues during this meeting. Importantly, \revision{researchers} did not interject unless a staff member directly asked for help using the tool. Detailed field notes captured all interactions with the new tool, such as common navigational pathways and points of friction.
As shown in Figure \ref{fig:Timeline}, Phase 3 then consisted of three additional observations, with staff using the final version of the new interface. 
The goal of these observations was to understand how staff were using the tool nearly eight months after \revision{the final version was} deployed. 

\subsection{Group Co-Design Sessions}
\label{sec:codesignSessions}
To facilitate an effective collaborative design process with staff, we led three group co-design sessions.
In preparation for the first co-design session, we generated discussion materials from a preliminary analysis of pre-deployment interviews and the first BRC meeting observation. Using an iterative open-coding process, we generated low-level codes from field notes such as ``reviewing bars all comes down to how well they write the log'' and ``BRC meetings can get off-track.'' We identified emergent patterns in the codes, leading to labelled clusters of codes. Examples of labels included ``difficulties with BRC workflow'' and ``staff have individual biases.'' 
After this preliminary thematic analysis, we transitioned to the \textit{Frameworking} and \textit{Generating Design Concepts} stages of the design ethnography framework, where preliminary research findings drive practical design decisions \cite{baskerville_design_2015}. These stages were enacted via a workshop on immersive ideation, and a feedback and brainstorming session with a working prototype.

The immersive ideation workshop took place after completing the preliminary analysis of interview and BRC observation data (see Figure \ref{fig:Timeline}). The goal of this workshop was to discuss our analysis with the staff, and embed these preliminary themes into the design of the new tool. The workshop began with an amended version of Holtzblatt's immersive \textit{wall-walk} exercise. \revision{For this exercise, the preliminary themes were projected digitally onto a blank wall as an affinity diagram. Participants contemplated these themes and wrote on physical sticky notes to add them to the projected diagram.} In accordance with Holtzblatt's method, participants were encouraged to ``think as designers'' as they creatively engaged with the design themes \cite{holtzblatt_10_2017}. \revision{Participants added a total of 30 new sticky notes to the wall, all containing additional considerations for each theme.} This exercise gave rise to a fruitful group discussion, resulting in new subthemes and practical design directions. As the group discussion continued, we directed the conversation towards concrete outcomes, paying special attention to possible navigation pathways and interactive features for the new interface. This led to three collaborative sketches which captured participants' design ideas. 

We developed an initial prototype to synthesize all ideas from this workshop and reflect the ideas back to staff in a working tool. This prototype was referred to Version 0 (V0), and was used as a \textit{technology probe} during the second co-design session: a one-hour group feedback and brainstorming session (see Figure \ref{fig:Timeline}). Hutchinson et al. \cite{hutchinson_technology_2003} define a technology probe as a technological artifact to help researchers understand their users in a real world setting, field-test technology, and inspire users to think creatively about new designs.
The initial working prototype met these aims \revision{during the feedback and brainstorming session}. As per the \textit{Generating Design Concepts} stage of the design ethnography framework \cite{baskerville_design_2015}, concrete design considerations arose from this collaborative free-form brainstorming discussion, where staff were encouraged to critique the prototype and contemplate how to improve it for live deployment.

A second brainstorming and feedback session took place part-way through the iterative deployment process. Similar to the first, a prototype of Version 2 of the interface, used as a technology probe, grounded creative design discussions to further crystallize staff needs. In contrast to the first feedback and brainstorming session, this session included real accounts of use cases from using the interface with live client data during BRC meetings. Therefore, the group had a far more concrete understanding of their own needs, \revision{resulting in} highly specific \revision{design} takeaways from this session.

\subsection{Iterative Deployment: Versions 1, 2, and 3}
The \textit{Prototyping} stage of the design ethnography framework \cite{baskerville_design_2015} began with the deployment of Version 1 (V1) -- the first version to be used during weekly BRC meetings with live updating client data -- and ended with Version 3 (V3). 
During each BRC meeting, the interface was projected onto the shared screen in the front of the room, and controlled by P05 throughout the discussion.

Participants gave \textit{in situ} feedback on V1 as they used it during BRC meetings. We then developed an early prototype of Version 2 (V2). Staff participated in a feedback and brainstorming session (section \ref{sec:codesignSessions}) centred on the early V2 prototype, allowing us to then finalize and \revision{deploy} V2. While the BRC used V2, they shared additional minor feedback via informal conversations and email correspondence. We refined V2 to create V3, and after deploying V3, we paused our involvement with the Drop-In Centre. The aim of eliminating our presence was for staff to use V3 for an extended period of time without feeling watched or having immediate access to a researcher. \revision{We reinitiated contact approximately eight months later to complete Phase 3, consisting of three final BRC meeting observations (section \ref{sec:BRCObservations}), and a round of post-deployment interviews (section \ref{sec:interviews}).} 

\subsection{Data Analysis}
According to Baskerville and Myers, 
``the researcher’s exit from the design setting(s) does not conclude the DE [design ethnography]. Rather, the researcher now needs to write up the findings and produce a DE'' (p. 36) \cite{baskerville_design_2015}. 
Thus, the final stage of this research aligned with the final stage of the design ethnography framework: \textit{Moving Out} \cite{baskerville_design_2015}. After completing the hands-on work of creating a\revision{n} interface for the BRC, it was now time to analyse all phases of this collaboration to generate transferable takeaways. 

To this end, we conducted a reflexive thematic analysis \cite{braun_reflecting_2019, braun_toward_2023, byrne_worked_2022} of all data: written notes from pre-deployment interviews, BRC meeting observations, the immersive ideation workshop, and feedback and brainstorming sessions, as well as audio transcripts from post-deployment interviews. In total, 29.5 hours worth of qualitative data was analysed. Throughout analysis, we aimed to identify themes with accompanying illustrative quotes that not only capture how staff experienced the new interface during BRC meetings, but also characterize the general relationship that staff have with administrative client data when making frontline decision\revision{s} about shelter clients.


The coding methodology was inductive \cite{byrne_worked_2022}; after \revision{becoming familiar with} the qualitative \revision{data} and taking notes of first impressions, we generated original codes by carrying out \revision{an extensive,} iterative open-coding \revision{process} \cite{braun_conceptual_2022, braun_toward_2023}. Our coding focused on capturing meaning as communicated by the participants (\textit{semantic} coding), rather than attempting to identify implicit meaning behind participants' words (\textit{latent} coding) \cite{byrne_worked_2022, braun_conceptual_2022}. 

Initial open codes included, for example, ``The best decisions come from conversations we have after we look at the data,'' and ``Arbitrarily deciding how long somebody should be barred based on one interaction is not fair.'' Throughout the coding process, we kept ongoing notes to document the iterative coding process, as well as reflections on how the data is being interpreted and any questions about the content of the data and potential emergent patterns. We then clustered codes into labeled categories. Examples of labels included ``Multifaceted and dissenting perspectives on the purpose of barring and BRC,'' and ``Data can flag parts of a client's story for future conversations.'' These categories brought analysis closer to delineating \revision{specific} factors in how staff interacted with the new tool during BRC, and how staff perceive the benefits and challenges of using data to make frontline decisions.  
Further iterative analysis resulted in 
a hierarchical structure of codes. For example, a parent category, such as ``Balancing Client and Staff Wellness,'' included several labeled child categories, which each included several low-level and \textit{in vivo} codes. In total, over 450 codes were generated from the data. All data was manually coded using NVivo 14 \cite{lumivero_nvivo_2024}. From this hierarchical structure, we generated a \revision{collection} of interrelated descriptive themes. The following sections describe the resulting landscape of themes.

\section{Co-Design Results: The Bar Review Data-Navigation Interface (BRDI)}
The final version of the new interface was called the Bar Review Data-navigation Interface (henceforth, ``BRDI'', \revision{pronounced ``birdy''}). BRDI addressed the drawbacks of staff members' existing interface, Guestbook. The practical design work of creating BRDI allowed us to delve into one specific decision-making context, \revision{the bar review committee, which revealed} the role that data can play at the Drop-In Centre. This section provides the first contribution of this research: a description of the newlt designed interface and an analysis of how participants used it to make difficult decisions about vulnerable clients. The design of BRDI is briefly described below, followed by a typical usage scenario, and interview themes that capture real-world experiences with BRDI.

\subsection{Design of BRDI}
\revision{Given the effort needed to read through dense administrative data, especially when logs can contain multiple paragraphs of text,} we expected staff to request some degree of data processing so that the tool would output abstracted metrics or data visualizations. We instead found that staff largely resisted \revision{abstracting away from the raw data, such as through bar counts or keyword summaries. Rather, they favoured interacting with unprocessed data that retains all granular detail needed for making high-stakes decisions about vulnerable people, such as written log notes.} After several iterations, our co-design resulted in a two-page tool. \revision{See Appendix \ref{appendixB} for annotated mockups (Figures \ref{fig:LookupPageUnfilt}, \ref{fig:LookupPageFilt}, and \ref{fig:DeepDive}) and anonymous screenshots (Figures \ref{fig:SSLookup} and \ref{fig:SSDeepDive}).}

The BRDI experience begins with the ``Lookup Page,'' which includes interactive elements to visually filter a list of barred clients. Once staff select a client, they can navigate to \revision{the client's} ``Deep Dive Page,'' showing all relevant data for discussing that client's bar. Co-design sessions and iterative feedback culminated in the following data needs for the Deep Dive Page: client's name, date of birth, age, type (shelter client or community client), last check-in date, housing program (if any), housing status (housed or not housed), personal health number, profile photo, all conditions of entry, and all restrictions of shelter services. The Deep Dive Page displayed this information within a visual hierarchy appropriate for BRC discussions. This page also prominently displayed all logs written about the client, along with their active and inactive bars, and a history of the client's check-ins in a column chart with barring history superimposed.


\subsection{BRDI Usage Scenario: Observations from the Field}
Every Thursday, P05 began the BRC meeting by launching BRDI and projecting it onto a wall for all to see. Using the Lookup Page, P05 would generate a list of clients who received Category 3 or 4 bars within the past seven days -- typically 20 to 40 clients. This list served as the meeting agenda. 

As P05 displays \revision{the first client's} Deep Dive Page, attendees read the projected client data, collaboratively interpret the data, and decide on an outcome they believe is fair for the barred client, the other shelter clients, and the staff. They may lift the bar, adjust its duration or category, replace it with a condition of entry, keep it unchanged, or seek further information from the people involved in the incident. 
If possible, the BRC will decide via consensus; otherwise, they will hold a vote. Once a decision is made, P05 navigates back to the Lookup Page to select the next client for discussion from the filtered list. They then ``Deep Dive'' on this next client, and continue this pattern until every client is discussed. \revision{Since BRDI is a read-only tool, P13 has Guestbook open on their own device to make any necessary changes to the database after each decision is made.}

BRDI's interaction flow merged smoothly with the flow of the meeting. Notably, the cyclical navigation between the Lookup and Deep Dive pages provided structure to discussions. \revision{Our analysis of the first BRC meeting observation, from before BRDI was deployed, highlighted the data-interaction obstacles staff faced when they used Guestbook to facilitate bar-review discussions.} Before BRDI's deployment, staff were required to navigate through the entirety of the Guestbook interface, in search of multiple scattered pieces of data to drive each point of discussion about every client. This process was laborious as it entailed sifting through 23 pages, all with aesthetically-identical flattened data tables, requiring manual text filters to narrow search results within each table. Given than BRC meetings are already long and emotionally tense, this navigation process in Guestbook was especially taxing. \revision{Further analysis revealed a plausible root cause of these issues: Guestbook does not cater to any one particular job function at the Drop-In Centre, and instead was designed to operate as a universal interface for all 300+ staff to use in several different contexts.} 
\revision{In contrast, BRDI is custom-built for a particular set of staff who make collaborative BRC decisions. This highly specialized design was essential in overcoming the aforementioned obstacles. Consequently,} BRDI allowed the BRC to follow a predictable sequence of navigations, creating a clear and orderly discussion framework for the meeting. 

\subsection{Experiences of BRDI During BRC Discussions: Interview Themes}
All seven post-deployment interview participants had favourable impressions of BRDI. P10 described it as ``\textit{perfect},'' ``\textit{powerful},'' and ``\textit{very clean and efficient.}''  P05 explained that BRDI ``\textit{makes it way more efficient [during BRC],}'' and that ``\textit{the Deep Dive function is amazing}.'' Other interviewees stated that BRDI ``\textit{helps with the flow of the meeting}'' [P06], and is ``\textit{very helpful [and] easy}'' [P09]. P12 explained that BRDI ``\textit{gives us an idea of how someone functions here, and it's very quick to see that, whereas in Guestbook it is not.}'' 

\revision{BRC members continued using BRDI after the university research collaboration officially concluded, although use slightly declined as new committee members joined the BRC. The researchers remain casually involved with the Drop-In Centre and have discussed holding workshops to train new staff and incorporate their feedback into future BRDI iterations. Additionally, this research sparked an ongoing ``dashboard'' project within the Drop-In Centre: the staff from this collaboration have initiated new conversations with the IT team to request improved tools to meet their various data needs, now that they have experience articulating these needs during our collaboration. According to P09, ``\textit{The dashboard right now, the [BRDI], is really cool. I would say it is a really good project that it started. ...It really set the standard higher.}''}

Three key themes capture participants' experiences with BRDI in comparison to their experiences with Guestbook: (i) BRDI is a ``one-stop shop'' that minimizes information overload, (ii) BRDI acts as a central reference point to support collaboration, and (iii) BRDI provides a holistic depiction of client circumstances amid conflicting staff perspectives. These themes pertain to using data when making collaborative high-stakes decisions about clients experiencing homelessness. Thus, the themes from this case study can serve as guidelines for future designers who aim to support staff in similar decision-making contexts. 

\subsubsection{A One-Stop Shop: Minimizing Information Overload}
Five of seven interviewees explained that BRDI consolidates all relevant information into a digestible form: ``\textit{We have the information on one page... it's like the one-stop shop}'' [P06]. This leads to a feeling of data-immediacy, where staff can see, ``\textit{right off the bat}'' [P13] the story the data paints about a client. This immediacy invites thoughtful critique of bar placements. P11 explained that the bar history chart allows quick identification of which bars to be sceptical of: ``\textit{With your visualization of it, [I can say] `Hey, this guy's got no history of this kind of behaviour, and all of a sudden he's got this Category 4 bar, a very severe bar, but he's got no history of bars. Like, what happened?'}'' P11 continued, ``\textit{...Whereas, if I didn't have that visualization [before BRDI], I would have just assumed, `Yep, he's got a bar. That's how it is, right?' But I'd have to intentionally look deeper if it's not in front of me}.'' 

In contrast, staff had not experienced this data-immediacy when they used Guestbook. \revision{Rather,} P05 explained that Guestbook caused feelings of ``\textit{information overload},'' where ``\textit{if I'm trying to get a full picture, it's easy for me to get lost in the weeds.}'' Importantly, though, having several data points together on one page can also risk information overload. When brainstorming additional BRDI features, P05 warned, ``\textit{If you're not careful, [BRDI] can turn into Guestbook... I think it'd be really easy to have too much information.}'' Fortunately, since BRDI was designed only for one particular data-usage context -- BRC discussions -- it could display all relevant data together while still avoiding feelings of information overload. 

\subsubsection{A Central Reference Point: Supporting Focused Collaboration}
While BRDI is projected, attendees also have their own devices open to \revision{take notes or} search for additional information to add to the conversation. In light of this dynamic, five of seven interviewees explained that BRDI acts as a central reference point to ground discussions. P05 explained, ``\textit{Everyone can see the information, so we'll know exactly who we're talking about [when using BRDI].}'' 
P13 said that BRDI helps with multitasking: ``\textit{It's helpful to have my screen, and [be] doing what I need to do, but also being able to look up at the log and read what happened... It definitely saves time}'' [P13]. Guestbook, in contrast, was designed for use by individuals, overlooking collaborative use-cases. Therefore, none of its pages clearly displayed general information about a client (name, photo, personal health number), simultaneously with historical details that depict that client's circumstances (logs, bars, check-in history). Instead, each datapoint was housed separately. 

Consequently, \revision{when P05 used Guestbook,} BRC members occasionally lost track of who they were discussing if P05 scrolled or navigated away on the projected screen: ``\textit{When someone is up there scrolling logs, and some people are still [looking] up here, some people are [looking] down there, it was a mess}'' [P06]. This would interrupt the flow of the meeting, so staff disengaged from the collaborative discussions. In contrast, BRDI ``\textit{helps with the flow of the meeting. It helps keep everyone on the same track. We're all getting the same information at the same time}'' [P06]. In summary, BRDI fosters the collaborative nature of BRC meetings by operating as a central reference point, creating common ground throughout each discussion. 

\subsubsection{A Holistic Depiction of Client Circumstances Amid Conflicting Perspectives}
A final theme emerged regarding the different roles BRC members adopt. 
Specifically, shelter staff approach discussions from a ``shelter perspective,'' meaning they focus on the short-term safety of all individuals in the building. Meanwhile, housing staff approach discussions from a ``housing perspective,'' prioritizing connecting clients to housing resources \revision{to meet their long-term goals}. P05 explained that having a variety of perspectives and departments represented in the BRC is crucial for healthy discussions. This way, all sides of the client's story are considered, and staff can ``\textit{check each other if we're not making the right decision}'' [P05]. Consequently, because they wear different ``hats'' [P06], the shelter team and housing team benefit from different features of BRDI. Specifically, the shelter team is ``\textit{looking for shelter check-ins, because that's the hat they wear. So they're looking for `When's the last time someone checked in?'}'' [P06], whereas the housing team is ``\textit{looking for what [housing] program they would fall under}'' [P06]. 

It was crucial that BRDI not neglect types of data only relevant to some teams. Otherwise, whole portions of a client's story could be pushed aside during discussions, \revision{sometimes due to conflicting values and implicit power dynamics between staff teams}. 
In Guestbook, for example, data relevant to the housing perspective 
were siloed away from data relevant to the shelter perspective. 
BRDI closed this gap by pulling disparate data together, creating holistic client narratives. With several data types shown on one page, staff could readily see information that would move the conversation forward and help them justify their arguments for or against various decisions. 

In conclusion, BRDI successfully addressed the BRC's needs. As mentioned in section \ref{sec:RelatedWork}, prior research on data tools in the homelessness sector focus on how ML-based tools can automate and expedite decision-making about vulnerable people. In contrast, BRDI did not automate decision-making, yet it still improved such processes by creating an optimal environment for staff to \revision{collaboratively interpret the raw administrative data. In P13's words, ``\textit{at the end of the day, ...the situations that we deal with are very complex. We’ll need to read everything in order to get the big picture.}''} Participants' hesitation to abstract away from the raw data \revision{prompts many lessons for the research community.} The following sections \revision{unpack this relationship between staff and data by detailing the perceived benefits and challenges of incorporating data systems into frontline work.}

\section{Findings: A Complex Relationship Between Staff and Data}
We now \revision{expand beyond the BRDI case study} to take a broader look at the relationship staff have with administrative shelter data. Overall, staff benefited from using data to make informed frontline decisions; however, these benefits were eclipsed by the challenges of holistically understanding nuanced client situations. Below, we present several interview findings related to these benefits and challenges, constituting the second contribution of this work. 



\subsection{Benefits of Using Data for Frontline Decisions}
\label{sec:Benefits}
Participants' appreciation of the Drop-In Centre's abundant administrative data can be understood in terms of three key themes. Administrative data (i) offer valuable insight into client circumstances; (ii) open lines of communication among staff and clients; and (iii) uphold fairness and accountability.

\subsubsection{Data Offer Valuable Insight into Client Circumstances}
When asked about their experiences with data, all interviewees shared examples of how administrative data illuminate the nuances of clients' stories. For example, if a client's data show recent check-ins after a significant gap, then the client may have unique feelings of ``\textit{guilt and shame [because] coming back to shelter is often worse than your first experience. ...Especially in [the] housing [team], data helps us understand patterns of behaviour that helps us understand what their needs are... and helps us shape what our programs do and how we serve people}'' [P12]. 

Check-in data over time can also reveal priorities for housing staff. P06, for instance, explained that the Drop-In Centre would benefit from a tool that triages clients into various internal housing streams \revision{based on building check-ins}. This hypothetical tool would take a client's check-in data as input, and output the most appropriate housing resource for them, such as ``diversion'' for new clients, case management for \revision{chronic shelter-users}, or other programs \revision{based on other shelter-use histories.} As P06 notes: ``\textit{If someone doesn't have any sleep check-ins, or it flags that they're new, it automatically goes to diversion. ...If you're a chronic shelter user and you have a certain number of drug poisonings or bars, you'd automatically be flagged for case management. ...If you're super episodic and we don't really see you that much, but you come in sometimes, we're likely to send you to the housing hub.}'' These comments exemplify that basic administrative data \revision{such as check-in histories} are influential for understanding clients' needs and ultimately developing strategies to \revision{help them overcome homelessness.}

Other simple administrative metrics can also raise flags about client behaviour. P05 explained that if a client has a high volume of building check-in data with minimal sleep program check-in data, then they might only be entering the building to deal substances to others. P09 also explained that if the client has received three or more bars for ``dealing,'' then they should be flagged for \textit{predatory} dealing -- repeatedly dealing to targeted individuals, possibly as part of an organized network. Additionally, P10 described that historic data, such as barring history and log history, can be foundational in creating threat-assessments of clients. Specifically, these metrics can reveal whether mental health complications could have caused the client's violent behaviour, or if their behaviour more likely resulted from conscious malicious intent. In summary, by incorporating administrative data into frontline workflows, patterns in client behaviour can become apparent \revision{to staff who know what to look for,} which can help guide ground-level program delivery.


\subsubsection{Data Open Lines of Communication Among Staff and Clients}
Six of seven interviewees explained that particular administrative data can open lines of communication among staff and clients. For instance, if bar entries are associated with clients who typically engage positively with the shelter, this can trigger crucial conversations with clients about their well-being and their needs: such \revision{outlier} data can be ``\textit{an opportunity for housing [teams] or health [teams] to follow up with this person when they can find them next, and try to have a conversation}'' [P13]. Patterns in check-in data similarly drive lines of inquiry about a client's well-being by sparking productive case-management conversations. P12 appreciates seeing \revision{temporal data patterns} like no check-in data in the beginning of each month. This allows them to ask the client ``\textit{`Where do you go in the first week of the month, every month?' And they're like, `Yeah, I blow my entire AISH [Assured Income for the Severely Handicapped] cheque on a hotel.' It's like, well, `Would you rather pay rent?' ...It kind of gives us a little bit of context about what they're up to, and opens up the lines of communication.}'' 

Another data pattern that sparks conversations is the overlap of a positive housing status with continued check-in data. P06 explained, ``\textit{If Guestbook says that you're housed, but you're still doing building check-ins two or three times a day, what are we missing?}'' They emphasized that technologies should alert staff of this pattern, describing that these situations ``\textit{would be a good time to ask [the client] questions, [such as] `Is it a food insecurity thing? Is it a roommate thing?' ...`Do you need a bed? Do you need a food bank referral?'}'' Staff also benefit from seeing data that classify clients as ``return-to-shelter'' cases. Staff emphasized that tools to isolate this client group would be helpful because ``\textit{trying to track the folks that come back after being housed is borderline impossible}'' [P12]. P06 explained that the context surrounding \textit{why} the client returned to shelter is not particularly important; instead, they would simply like to identify this subset of clients and flag them for follow-up conversations. 


Data patterns can also trigger conversations \textit{between staff}, about clients. For example, if there is a disproportionately high number of entries about one client from one staff member, then managers could investigate potential larger issues. P11 explained: ``\textit{[If] there's seven logs about the one individual from one staff member... then we would explore other options and look at health services and say, `Do they \revision{[the client]} have fixation issues? Are there stalking issues happening? What's going on here?'}'' Overall, when frontline staff regularly engage with administrative data, they become sensitive to the patterns that precipitate deeper investigations into a client's behaviour, opening lines of communication with clients and sparking constructive discussions between staff.

\subsubsection{Data Uphold Fairness and Accountability}
Finally, four of seven interviewees explained that administrative data provide a means for staff to justify their decisions, while being held accountable to clients and to each other. For instance, when the BRC decides to revise a bar, the person who originally placed the bar may want a justification. The first time this happened to P11, they ``\textit{took it very personally},'' believing their original bar placement was \revision{warranted}. After a BRC member showed them specific data points that led to the decision, P11 understood the change. This illustrates how data can maintain accountability of the BRC \revision{to other staff.} Additionally, staff opinions about clients may differ if they only see clients during particular times of day. Amid these biased viewpoints, the data can provide a full ``\textit{24/7}'' [P09] account of shelter events, providing staff a means to justify bar placements to others who may lack understanding of the relevant details.

While the above examples pertain to staff being held accountable to each other, administrative data also allow staff to be held accountable to their clients. For instance, when clients check in, staff \revision{look for any active bars, along with} qualitative notes describing the barring incident. P11 explained that they need to refer to this qualitative data at the front entrance to justify to clients why they were barred, and ensure it is fair: ``\textit{If they are barred, I want to know why they are barred. Maybe I need to have a conversation with them about what bar happened. Maybe I could influence their behaviour in a certain way. But it would give me enough context as to tell them `You are not allowed in, and here's the reason why.'}'' P10 additionally emphasized that ``\textit{we need to make sure that we're respectful towards the clients. They're humans at the end of the day [and] this is one of the few places that they can come to and feel safe and [get] food and services. So, we don't want to keep them from that if we can't prove that they did anything wrong.}'' Since staff are ``\textit{in a position of respect and trust}'' [P10] they must always provide ``\textit{unbiased documentation... to provide that evidence to our peers, our superiors, our partners, et cetera}'' [P10].
In general, participants feel that administrative shelter data can ground decision-making in documented records of events, providing the means for staff to be held accountable not only to each other, but also to their clients.

\subsection{Challenges to Using Data when Making Frontline Decisions}
\label{sec:Challenges}
While staff appreciate the above benefits, they also face challenges when using data for frontline decision-making. 
Presented below are three key challenges: (i) data are not clear due to bias and lack of standardization; (ii) data are incomplete because they overlook nuance and decontextualize client behaviour; and (iii) relying on data can conflict with frontline values. 

\subsubsection{Data are Not Clear: Bias and Lack of Standardization}
All staff stated that administrative data contain bias and are not standardized. For example, log data become messy when ``\textit{everyone has their own writing style}'' [P10]; while some staff are ``\textit{very meticulous},'' others will document only ``\textit{things that they think are important}'' [P05]. Log data can additionally have ``\textit{emotions attached}'' [P06]. P06 explained that, rather than having neutral documentation that describes, for example, ``\textit{at this time this happened, and then this happened, and I responded this way, and this was the outcome},'' some documentation will be more akin to ```\textit{then they called me this, and then they were like... belligerent.' You're like, `Okay, well, what does that mean?'}'' [P06]. Varying writing styles and vague, emotive terms, \revision{such as `belligerent,'} can compromise the clarity of log notes, making them unreliable as grounds for high-stakes decisions.



P13 also pointed to oversimplified logs that are insufficient to justify \revision{corresponding} bars. They described a barring incident where the documentation stated ``\textit{This person was unhappy at me and they spit at me.}'' P13 explained, ``\textit{You need a little bit more detail to put in that bar. Did they just spit at you? Where did the spit land? Did it land on you? Did they spit }at\textit{ you but it didn't connect?}'' These comments also underscore discrepencies in staff focus when documenting incidents: some attend to the client's \textit{intentions} (the client intended to spit at the writer), while others attend to the \textit{outcome} of the client's behaviour (the writer was not actually spat on). This discrepancy was echoed by P11 when describing a bar placed with the explanation, ``\textit{Harm to Staff},'' and the accompanying documentation stating ``\textit{The client `attempted to harm the staff.'}'' P11 described their reaction: ``\textit{So, you weren't harmed, then? ...In my view, I'm like, `How can you justify such a bar?'''} These comments demonstrate not only that qualitative log data are unclear, but that quantitative bar-placement data can \revision{consequently become} biased as staff continue to place bars for inconsistent reasons.
In summary, data are not clear because of the following issues: documentation contains varying degrees of specificity, some \revision{qualitative} data includes subjective interpretations and vague emotive language, and staff may tend to focus on different aspects of each incident.

\subsubsection{Data are Incomplete: Overlooking Nuance and Decontextualizing Client Behaviour}
Regardless of bias or lack of standardization, five of seven interviewees explained that the \revision{administrative} database is incomplete because it fails to capture all aspects of client behaviour. 
This was in part because strength-based attributes of client circumstances are often neglected in documentation. Although staff would appreciate positive, strength-based data for a well-rounded depiction of the client, they more readily document negative interactions that necessitate an actionable response.

\revision{Beyond missing strength-based documentation,} data do not fully reflect the complex culture of clients at the Drop-In Centre, such as gang affiliations and interpersonal client dynamics, creating more gaps in the database. For example, P12 shared the details of a chaotic incident where multiple people attacked one person. A woman gave a knife to a man, who then stabbed another client. P12 pointed out the many pitfalls of relying on administrative data when reviewing these multilayered incidents. They explained that it is ``\textit{very difficult}'' to determine client motives and coercive factors: ``\textit{We have to put ourselves in the shoes of the man who took the knife and stabbed the person. ...What are the risks of him }not\textit{ stabbing the person? What kind of control does this woman have over him? ...Is she affiliated with gangs? And if he doesn't stab this person, is he at risk?}'' P12's comments highlight the obscure power dynamics between clients, causing ``\textit{the risk of not doing things sometimes [to be] huge.}'' These complexities are \revision{not reflected} in the data that \revision{staff use to make decisions about clients.} 

As well, administrative data fail to recognize the uniquely stressful nature of the shelter environment, where clients are in constant ``\textit{survival mode}'' [P06], and their behaviour is ``\textit{different than how it would be in their own home or out in community}'' [P06]. Moreover, data can only present events in isolation, devoid of the shelter context. P06 articulated that bar placements with the accompanying description, ``failure to de-escalate,'' are perhaps unjustified: ``\textit{If someone is yelling and screaming, but they're not hitting anyone and they're just having an outburst, to me, I'm like, this is actually great, because no one got hurt. They got to have their reaction, and that's fair, given what it's like living in shelter. ...We bar normal reactions and then hold people to, everyone to, the same standard of behaviour. ...That's, I think, where we fail people}.'' Since the documentation is stripped of this context, staff must use their own discretion to add nuance to the data and recontextualize client behaviour. Overall, these inherent weaknesses in administrative data cause issues when staff attempt to rigidly follow protocols and base their decisions \revision{solely} on what the data depict. In summary, three aspects account for why the data are incomplete: positive client interactions are typically not recorded, client culture and interpersonal power dynamics are not reflected, and events are documented without taking the shelter context into account.



\subsubsection{Relying on Data Can Conflict with Frontline Values}
Since the data are not clear nor complete, four of seven interviewees explained that they avoid relying on data, making sure to inject ``\textit{humanness}'' [P06] into their decision-making. For example, P05 felt it was important to be aware of subjective bias when making BRC decisions. Participants further explained that the ``\textit{purpose of BRC}'' is to navigate the ``\textit{grey}''  [P09], and that ``\textit{we do our best work in the grey area}'' [P12]. 
The ``grey'' in these comments refers to instances when staff realize that data cannot wholly represent barring incidents; thus, they refrain from strictly following a set of prescribed decision-making heuristics based on this data.
Instead, the BRC must interpret the data and collaboratively recontextualize it to make fair decisions.

P06 emphasized the role that data and technology play during bar-review decisions: ``\textit{Once we have all looked at the data and we all have an understanding of what the event was, then we can turn to each other and say, `Paint the picture of more of who the person is, and what the person needs.'}'' P06 elaborated: ``\textit{The technology is the canvas, and then we kind of do the painting... I think our best decisions come from good conversations after we have the canvas. ...The data is so important because it captures what we need to talk about, but I think the outcomes are coming from the humanness in the conversation.}'' Thus, data \revision{systems} form an effective framework for decision-making; however, staff are also careful to not solely rely on data when making BRC decisions. 

P10 similarly found it helpful when BRC members add personal anecdotes about the client to discussions. They believe that adding this subjectivity, although external to \revision{what is presented in} the data, invites staff to approach decisions with more compassion.
These comments highlight the importance of confronting the limits of using data in this context. Lastly, P06 explained the risks of over-relying on data to understand a client: ``\textit{It's easy when you have a lot of bars to get labelled as someone who's dangerous. Or you get a reputation in shelter.}'' They further perceived a risk of \revision{dehumanizing} clients, where ``\textit{If we relied on just the data, we would miss a lot of the information that [shows] the whole person.} \textit{...I don't want anyone to forget that that's actually someone who's a person and who has a family and, you know, they're probably very missed. ...It would be a huge fault if we just relied on what the data was telling us, and not the `human' approach to it}'' [P06]. These excerpts all emphasize that staff navigate a delicate balance between rigidly following protocols by relying strictly on recorded data, versus using their professional discretion to acknowledge the limits of data, and sometimes look beyond it. 

\revision{In conclusion, all seven interviewees described ways in which administrative data offer insight into client circumstances, while all also pointed to how these data are limited due to bias and variability. Six of seven interviewees explained that data patterns can open lines of communication, and four explained that using data in their work ensures fairness and accountability. Meanwhile, five interviewees believed data can overlook important nuances in client circumstances, and four felt that relying on data can conflict with frontline values.}

\revision{
\subsection{A Continuum of Reliance on Data Systems}}
\revision{In addition to the above benefits and challenges, interviews foregrounded varying degrees of willingness to let data systems drive decision-making. Whereas our collaboration focused on decisions made by the BRC, interviewees also mentioned several other frontline decision-making scenarios at the shelter. For example, P06 described how they triage clients into appropriate housing programs, and P13 described decisions to reach out to clients with unique needs who have returned to the shelter after previously being housed. For every new decision, staff felt differently about relying on data systems to guide their decision-making. We interpret this as a continuum between reluctance and willingness to outsource parts of their decisions to data systems, which we call the \textit{data-outsourcing continuum}.}


\revision{Most prominent in this case study was the BRC decision-making scenario.} In this scenario, over-relying on data posed significant risks to making equitable decisions about vulnerable populations. Staff here are \textit{reluctant} to outsource their decisions to data \revision{systems}. Instead, they prefer to take ownership of the decision-making process, with client data acting only as a starting point for collaborative discussions where they can then offer subjective interpretations of client circumstances. \revision{Housing program triage, as described by P06, exemplifies a scenario more characterized by a willingness to rely on data systems for decision making.} Importantly, these triage decisions do not dictate the allocation of actual housing to clients, but instead determine which internal case-management resources are \revision{offered} to which client. For these triage decisions, staff are comfortable relying exclusively on administrative data, such as check-in patterns and other historical records to make these decisions. They trust that these decisions can be entirely \revision{``}outsourced\revision{''} to a data \revision{system}, without human input or critical discussion of what is depicted in the data. 

\revision{
\section{Discussion: Implications for Design and Future Work}}
\label{sec:DOC}

The previous sections unpacked \revision{tensions between the} benefits and challenges that frontline staff faced when using data to make collaborative, high-stakes decisions about vulnerable people experiencing homelessness. \revision{Overall, staff are grateful for the extensive data infrastructure at the Drop-In Centre, and make efforts to use this data in their daily work. When staff ground their frontline decisions in administrative shelter data, they become attuned to patterns in the data that reveal hidden details about clients' stories, and these patterns can trigger crucial conversations among staff and clients. Some staff believe the data can act as a universal source of truth to be held accountable to. However, when making life-altering, time-sensitive decisions about clients, such as whether or not to allow them into the building after misconducts, staff grapple with the pitfalls of being entirely ``data-driven.'' Although staff strive to base their decisions in their organization's administrative data systems, especially when they have so much data available to them, they are nonetheless acutely aware of data bias and variability, and believe essential context is missing from these systems. Ultimately, some staff perceive data usage to be at odds with the human-centred values that are at the core of shelter work.}

Prior literature highlights that staff in the homelessness sector have resisted using technologies that automate decisions about vulnerable clients. \revision{The present study} provides a framework for possibly understanding why these prior computational tools were not well-received. For instance, the VI-SPDAT \cite{orgcode_message_2020} was created to assign quantitative scores to clients' levels of vulnerability, for the purpose of prioritizing who should receive housing. \revision{These decisions have huge impacts on clients' lives, and } 
staff are aware that the \revision{original} data used for these vulnerability scores are often confounded by several biases \cite{slota_feeling_2022}. Rather than \revision{willingness to ``outsource'' to data}, Slota et al. \cite{slota_feeling_2022} discussed a tangentially related concept: ``a feeling for the data.'' They explained that ``\textit{having `a feeling for the data' means having a nuanced understanding and appreciation for what the data is, where it came from, and its strengths and limitations}'' (p. 727) \cite{slota_feeling_2022}. Our work has extended this concept by further unpacking staff members' understanding of, or \textit{feeling for}, the strengths and weaknesses of their data. With their unique ``feeling'' for the data, Drop-In Centre staff easily identified the role that data can and cannot play within any particular decision-making scenario, giving rise to their willingness or reluctance to rely on data and data-driven tools. As such, technology designers must realize that, as outsiders, they cannot so quickly assume what level of data-outsourcing is acceptable. Rather, it takes time to develop a \revision{sufficiently calibrated} feeling for the data, similar to that of frontline staff. 




When designers aim to make new supportive technologies for human-centred settings, the data-outsourcing continuum offers a useful lens through which to understand users' readiness to outsource \revision{parts of their decisions} to data. Based on this paradigm, designers can ask better questions about decision-making tendencies. For example, designers can consider the following questions before designing new tools: 
\begin{itemize}
\item Do decision makers trust the accuracy and completeness of available data enough to make decisions based exclusively on that data?  
\item How much room should a decision-aid technology leave for staff to \revision{subjectively} interpret the data before making final decisions? 
\item What elements of the decision-making process can be reduced to a rigid set of heuristics, and what elements cannot? 
\end{itemize}
It could be beneficial for HCI researchers to collaborate with those in psychology to create a standardized scale to operationalize staff willingness or reluctance to outsource various decision-making processes to data. The development of such a scale would include questions similar to those listed above, and the results could guide designers in understanding what place computational tools may or may not have in each decision-making process.

\begin{figure}[h]
    \centering
    \includegraphics[width=1\linewidth]{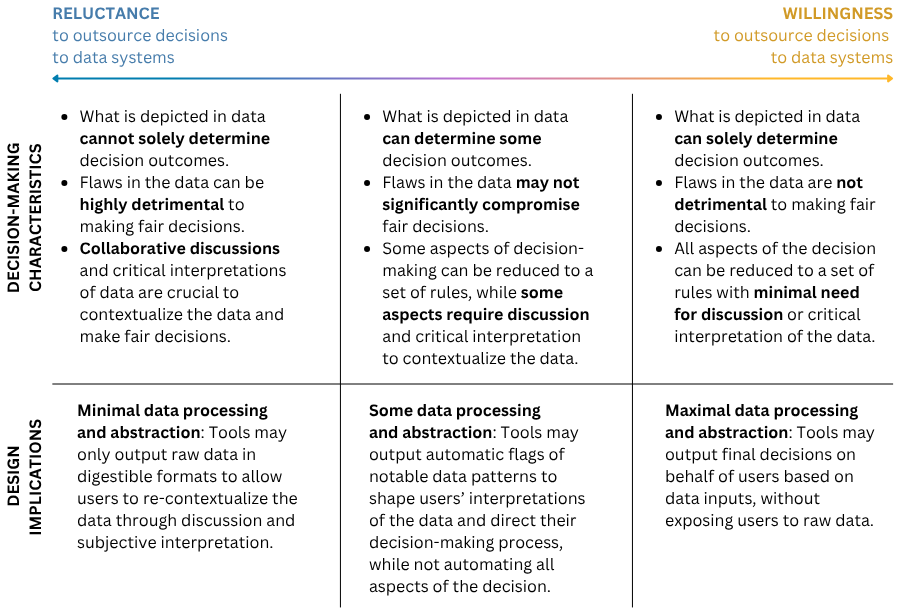}
    \caption{Summary of decision-making characteristics and design implications based on various degrees of reluctance or willingness to outsource decisions to data systems, for designers to consider before creating data-driven decision-aid tools. }
    \Description{A matrix diagram with three columns and two rows. The columns sit along a spectrum, where the leftmost column is labelled "reluctance to outsource decisions to data systems" and the rightmost column is labelled "willingness to outsource decisions to data systems." The top row is labelled "decision-making characteristics" and the bottom row is labelled "design implications."}
    \label{fig:DOCSummary}
\end{figure}

Based on \revision{users' willingness or reluctance,} designers could then consider the degree to which their technology should ``take over'' the decision-making process, as summarized in Figure \ref{fig:DOCSummary}. If a scenario \revision{precipitates much \textit{reluctance} to data-outsourcing} (leftmost column of Figure \ref{fig:DOCSummary}), then the technology should merely display raw data to decision makers and leave ample room for human interpretation \revision{of what story the data tells about a person}. \revision{Like in the case of BRC decisions,} designers can allow decision makers to recontextualize and ascribe meaning to the raw data, simply by bringing relevant data together \revision{and presenting them in an accessible interface.} Alternatively, if the use-case is towards the \textit{centre} of the data-outsourcing continuum (middle column of Figure \ref{fig:DOCSummary}), then decision makers may benefit from technology that performs a small degree of data processing, perhaps by generating flags or relevant insights about notable patterns in the data. The technology, then, bears the responsibility of finding and presenting unusual data patterns to staff that may prompt further action. This could be helpful to provide some direction to the decision-making process, while still not entirely dominating it. Finally, if the \revision{decision lends itself to being easily outsourced to data systems} (rightmost column of Figure \ref{fig:DOCSummary}), then it may be beneficial for designers to create ML-based or other fully-automated tools that take client data as input, and output final decisions for staff to act on. 

\revision{Overall, our fieldwork shone light on the realities of making ``data-driven'' decisions when dealing with the lives of individuals in vulnerable positions. The conclusions of this work are likely transferable to other care-work domains where staff help vulnerable individuals navigate life-threatening circumstances. For instance, the issues foregrounded in this work echo challenges faced by asylum workers. Specifically, prior work warns against ``datafying'' decisions around granting asylum to refugees, explaining that the rising popularity of data science can cause critical questions to be ignored about the way displaced individuals are construed as non-credible applicants by incomplete datasets \cite{rask_nielsen_data_2022, kaltenhauser_beyond_2025}. The overlooked flaws in initial datasets can then cause cascading issues for ML decision-aid tools in care work \cite{rask_nielsen_data_2022, moon_datafication_2025}. This apprehension around datafication was also found in our study, inviting similar scepticism about how useful ML-based decision-aid tools can be in the homelessness setting. However, further investigation (similar to \cite{kuo_understanding_2023}) is still needed to uncover the exact reasons for this apprehension, and how it translates to perceptions of ML tools.}

\revision{Notably, in other domains such as healthcare, individuals also make high-stakes, life-altering decisions about patients, but do not exhibit such hesitation to ground their decisions in data systems. Therefore, our conclusions are not necessarily transferable to all high-stakes decision-making contexts. Thus, this research raises questions about why decision makers hesitate to use data in the homelessness context specifically. A possible root cause is low data quality due to limited resources for data upkeep in the nonprofit sector \cite{voida_homebrew_2011}. Although, a key factor may be that shelter work is heavily contingent on the personal relationships staff build with clients -- relationships which can be harmed if staff prioritize efficient documentation over providing face-to-face care \cite{tracey_recordkeeping_2023}, or favour data representations of clients' behaviour over in-person interactions \cite{karusala_street-level_2019}.}

\revision{Our} study also raises questions about what factors influence individuals' \revision{willingness to outsource to data systems, given each particular decision-making context.} Based on the scenarios participants discussed in this study, we suggest that the \textit{type} of data that \revision{underpins} the decision-making process is a key factor in \revision{decision makers' willingness to ground their decisions in that data}. For example, narrative, qualitative data -- such as that used in BRC discussions -- \revision{tend to necessitate much critical interpretation}. Meanwhile, quantitative metrics like binary check-in data -- such as that used in the housing program triage scenarios -- \revision{may allow for greater automation}. 
Furthermore, the degree to which each decision is considered ``high stakes'' might also influence willingness to outsource to data systems. For example, bar-review decisions have direct impacts on the lives of clients, and while internal housing program triaging decisions can also change the trajectory of a client's recovery, their influence is less dire. It is important to also note that willingness and reluctance to base decisions on data varied slightly between participants within the individual BRC context. \revision{Housing team members who focused on the long-term trajectory of clients' lives tended to encourage more critical interpretation of what picture the data depicted of each client. In contrast, security staff and those focused on the immediate safety of the shelter were more willing to follow a protocol based solely on what is shown in the data.} Thus, future work could pay particular attention to \revision{conflicting values and potential power dynamics between stakeholders, to determine if our conclusions are further nuanced by these dynamics}. This work also invites exploration into how stable or dynamic each decision is, with respect to its placement on such a data-outsourcing continuum. It is possible that willingness to outsource decisions to data would change based on the quality of the data (continuum placements could be \textit{dynamic}); alternatively, continuum placements may be \textit{stable}, in that certain scenarios may \textit{always} necessitate a high degree of critical interpretation and human judgment, regardless of data quality.

\section{Conclusion}
This work was motivated by the need for more specialized technology to support frontline staff at emergency shelters and other resource-constrained community-based nonprofits. Most work in this area employs methods such as interview studies that place the researcher at a careful distance from domain experts. While such methods are important for understanding users' attitudes, it is crucial for the HCI and CSCW communities to also conduct real-world co-design-based inquiries to gain detailed, experiential knowledge of how to support users. By leading a co-design project with the Calgary Drop-In Centre, we generated several insights regarding how frontline staff use administrative data when making high-stakes decisions about vulnerable clients.
Hands-on work within the bar-review scenario revealed staff members' hesitance to rely solely on data \revision{systems} to make decisions, due to fear of overlooking nuances of each client's story. Instead of using computational tools that would abstract away from data, they preferred to sift through the raw data to ensure they personally interpreted each piece of information. Discussions with staff about other scenarios then brought to light the presence of a \textit{data-outsourcing continuum}, where this reluctance changed based on the features of each decision-making scenario. This continuum serves as a \revision{paradigm} for designers to use when approaching new design contexts and building data-driven, decision-aid technologies. Before building new tools, designers may reflect on their assumptions about the role data \revision{can} play in their \revision{particular} design context. After assessing users' willingness to outsource to data, designers may then create decision-aid tools that align with human-centred values, and leave appropriate room for critical interpretation of data systems.

\begin{acks}
The authors would like to gratefully acknowledge the support of Making the Shift, the Calgary Drop-In Centre, and the Government of Alberta. This study is based in part on data provided by Alberta Seniors, Community and Social Services. The interpretation and conclusions contained herein are those of the researchers and do not necessarily represent the views of the Government of Alberta. Neither the Government of Alberta nor Alberta Seniors, Community and Social Services express any opinion related to this study. Thank you as well to the staff at the Calgary Drop-In Centre who participated in this research, to Donna-Lee Wybert for her support in editing an early version of this manuscript, and to the anonymous CSCW reviewers for their thoughtful and constructive feedback. 
\end{acks}
\clearpage
\bibliographystyle{ACM-Reference-Format}
\bibliography{bibliography}



\appendix
\clearpage
\section{One-on-One, Semi-Structure Interview Guides}
\label{appendixA}
\revision{This appendix contains the interview guides for both rounds of one-on-one, semi-structured interviews: pre-deployment interviews and post-deployment interviews.}

\begin{figure}[h]
    \centering
    \includegraphics[width=0.95\linewidth]{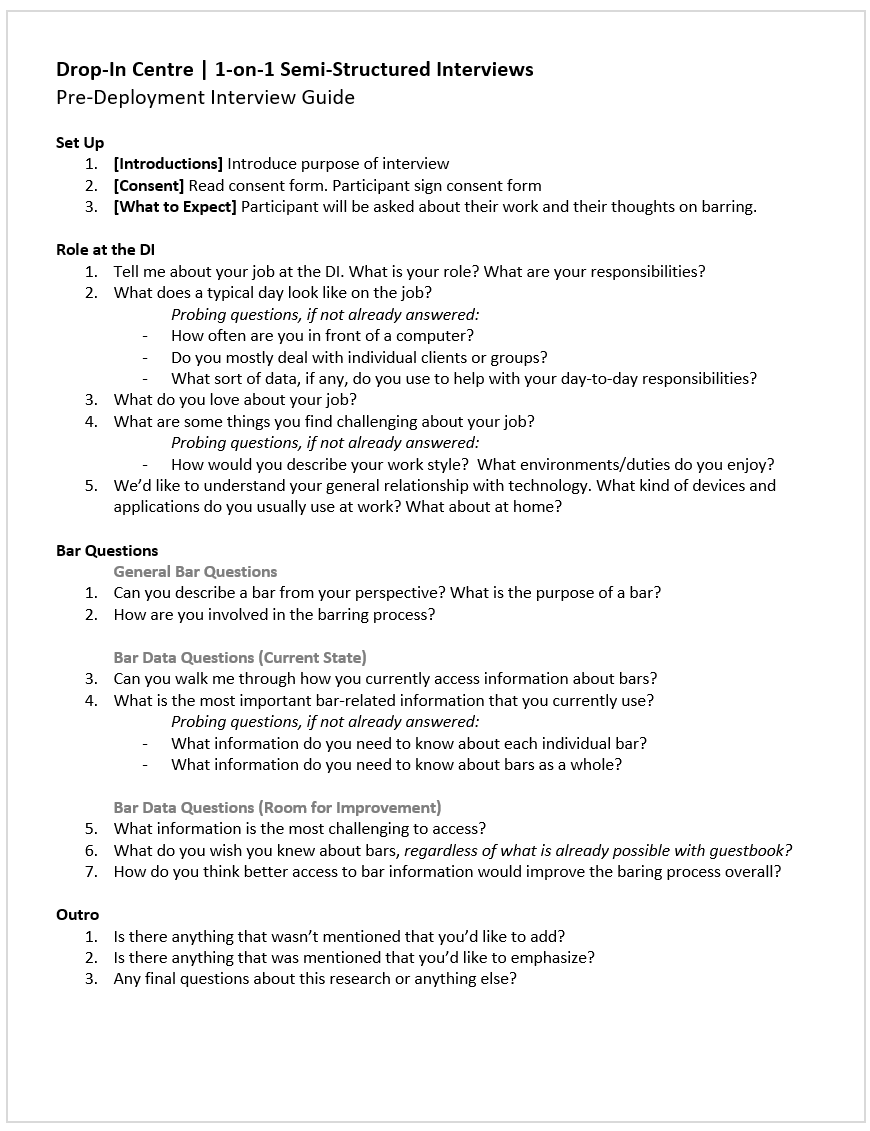}
    \Description{A screenshot of the pre-deployment interview guide, including interview setup, a section of questions titled "role at the DI", and section of questions titled "bar questions", and an interview outro.}
    \label{fig:PreDeploymentInterviewGuide}
\end{figure}

\begin{figure}[h]
    \centering
    \includegraphics[width=0.95\linewidth]{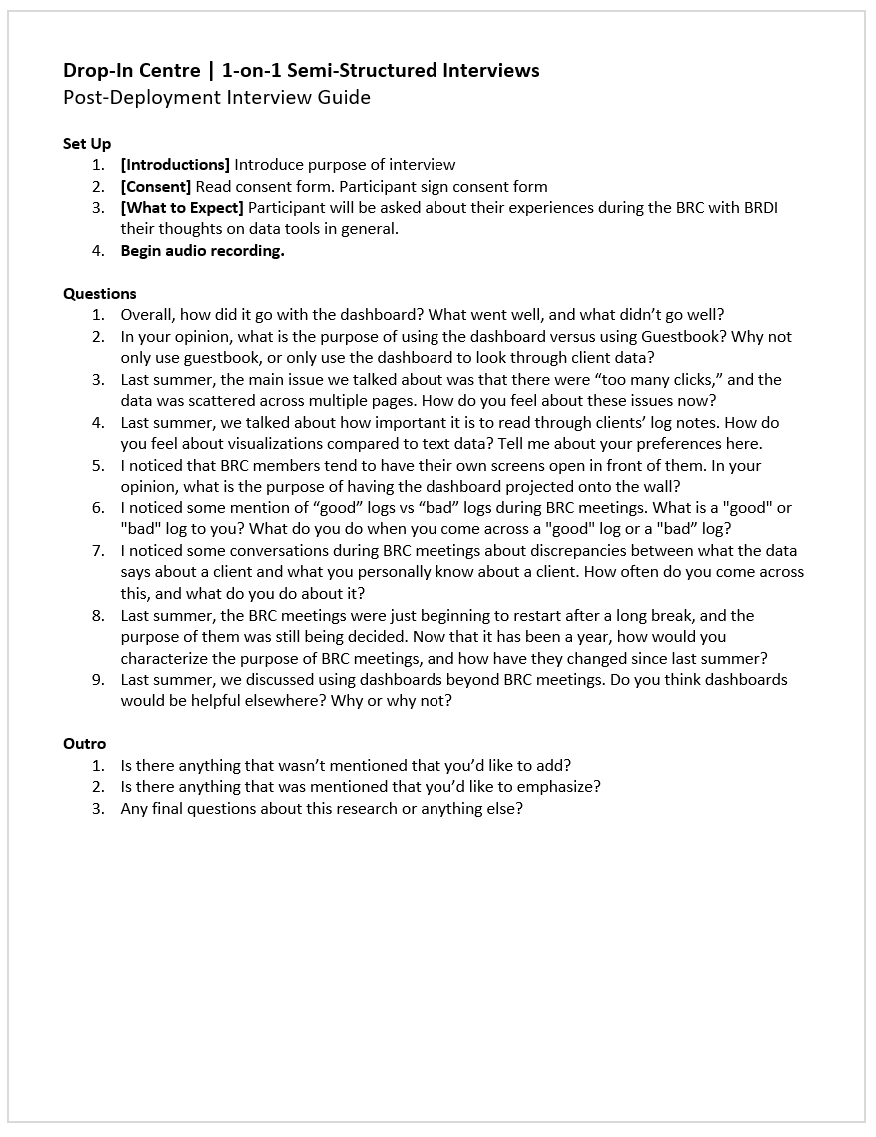}
    \Description{A screenshot of the post-deployment interview guide, including interview setup, all questions, and an interview outro.}
    \label{fig:PostDeploymentInterviewGuide}
\end{figure}

\clearpage
\section{BRDI Mockups and Screenshots}
\label{appendixB}
\begin{figure}[h]
    \centering
    \includegraphics[width=1\linewidth]{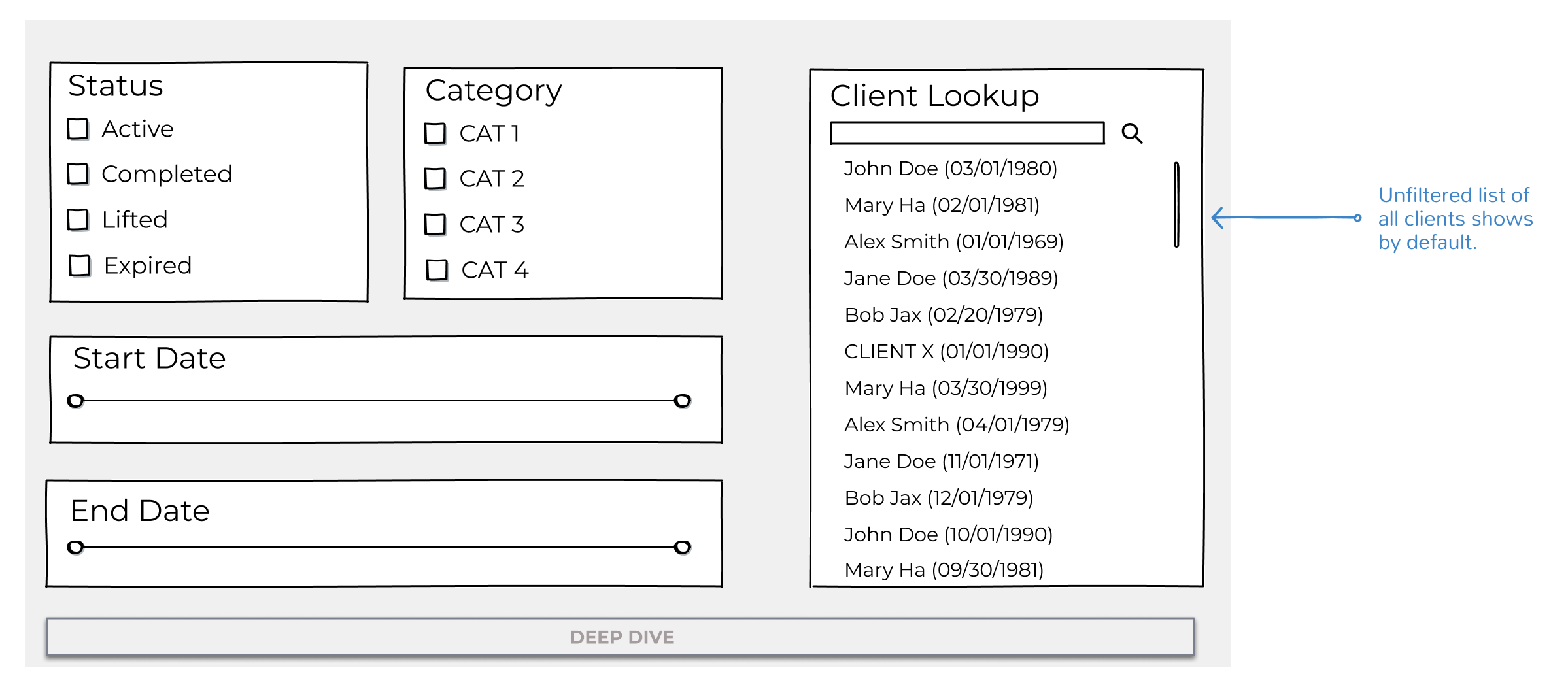}
    \caption{BRDI Mockup: The Lookup Page, before filtering the client list.}
    \Description{A wireframe style mockup of a screen with five panels. One is titled "client lookup" and is a list of client names. Annotations explain that the list is filtered based on the bars that each client's data contains. The other four panels are filters for the client lookup list: one filters based on bar status (active completed, lifted, and expired), one is for bar category (CAT 1, CAT 2, CAT 3, and CAT 4), and there are sliders for bar start dates and end dates. No filters are currently selected, and a "Deep Dive" button at the bottom of the mockup is greyed out.}
    \label{fig:LookupPageUnfilt}
\end{figure}

\begin{figure}[h]
    \centering
    \includegraphics[width=1\linewidth]{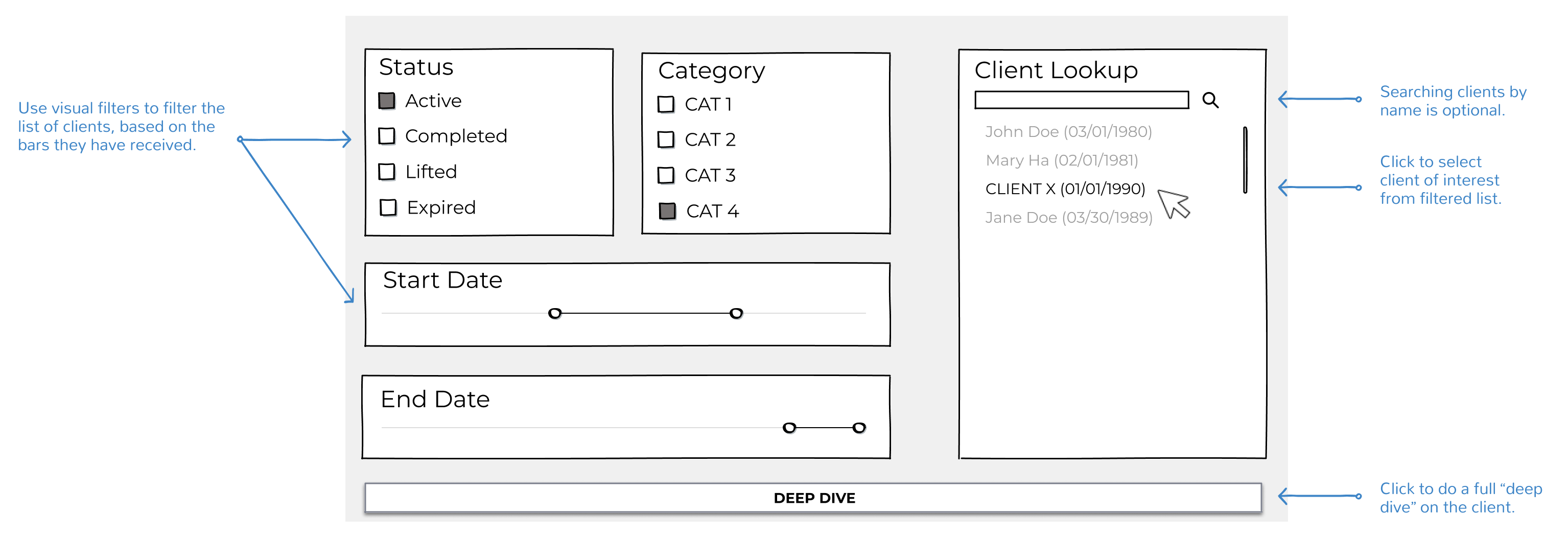}
    \caption{BRDI Mockup: The Lookup Page, after filtering the client list and selecting a client.}
    \Description{The same wireframe style mockup as Figure 3, now with options in each filter selected, and the client name list narrowed down to only display four names. One client, "Client X", is being selected with a cursor icon. The "Deep Dive" button at the bottom of the mockup is now active.}
    \label{fig:LookupPageFilt}
\end{figure}

\begin{figure}[h]
    \centering
    \includegraphics[width=1\linewidth]{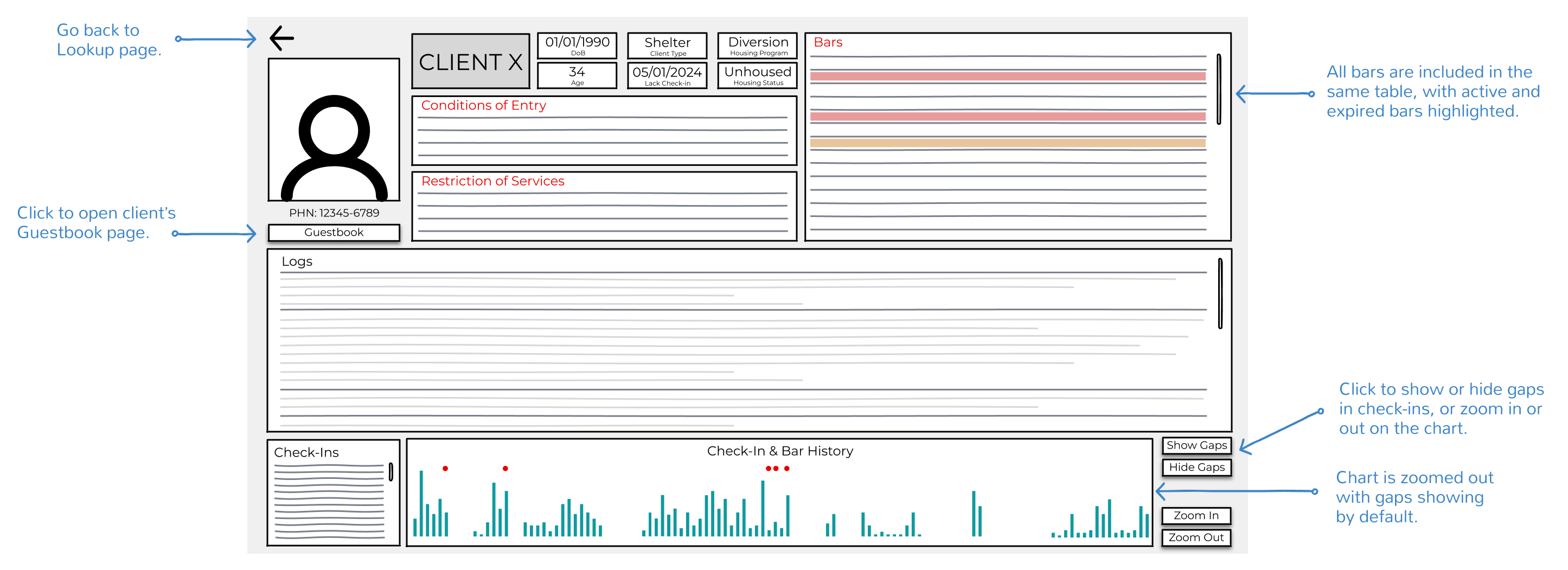}
    \caption{BRDI Mockup: The Deep Dive Page for one particular client.}
    \Description{A wireframe style mockup of the "deep-dive page" for Client X. The screen contains several panels, including a profile picture with the client's PHN and a button to go to their Guestbook page; the client's name, date of birth, age, client type, housing program, last check-in date, and housing status; a history of bars panel with active bars highlighted in red and expired bars highlighted in orange; a conditions of entry panel; and a restriction of services panel. The largest panel in the middle of the screen displays all logs written for this hypothetical client (text in this mockup is represented with horizontal lines), and on the bottom of the screen is a column chart representing the client's building checkin history each day, with small red circles on some days indicating hypothetical barring incidents.}
    \label{fig:DeepDive}
\end{figure}

\begin{figure}[h]
    \centering
    \includegraphics[width=1\linewidth]{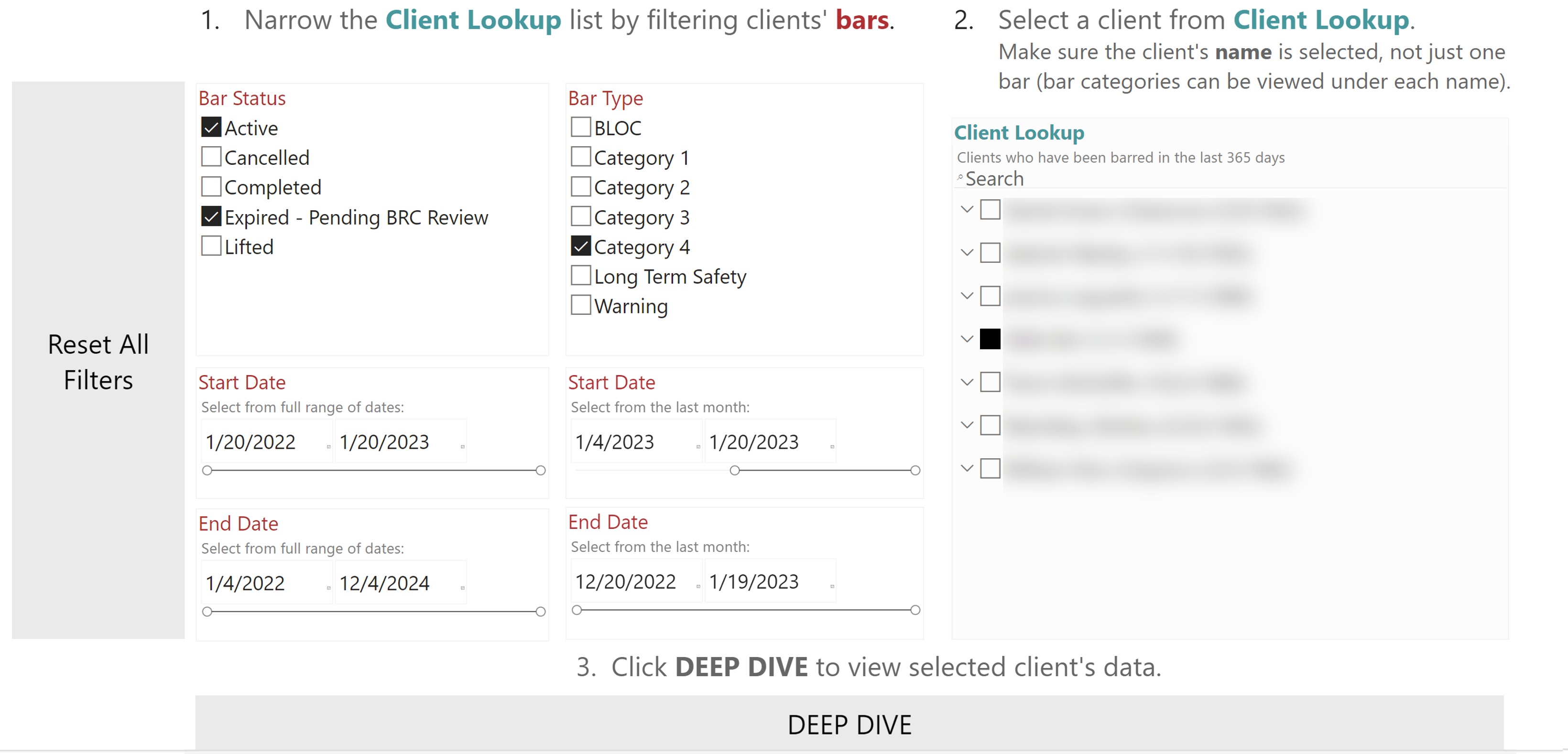}
    \caption{BRDI Screenshot: The Lookup Page, after filtering the list of clients based on their bars, and selecting one to conduct a ``deep dive'' on.}
    \Description{A screenshot of the BRDI's lookup page, similar to the mockups in Figures 3 and 4. There are bar status and bar type filtering panels, as well as start date and end date sliders, and a "Reset all filters" button. The filters narrow the client lookup list which displays blurred client names. A "deep dive" button is shown on the bottom of the screenshot. The screenshot additionally includes 3 instructions for the user: "1 . Narrow the client lookup list by filtering clients' bars", "2. Select a client from Client Lookup", and "3. Click Deep Dive to view selected client's data."}
    \label{fig:SSLookup}
\end{figure}

\begin{figure}[h]
    \centering
    \includegraphics[width=1\linewidth]{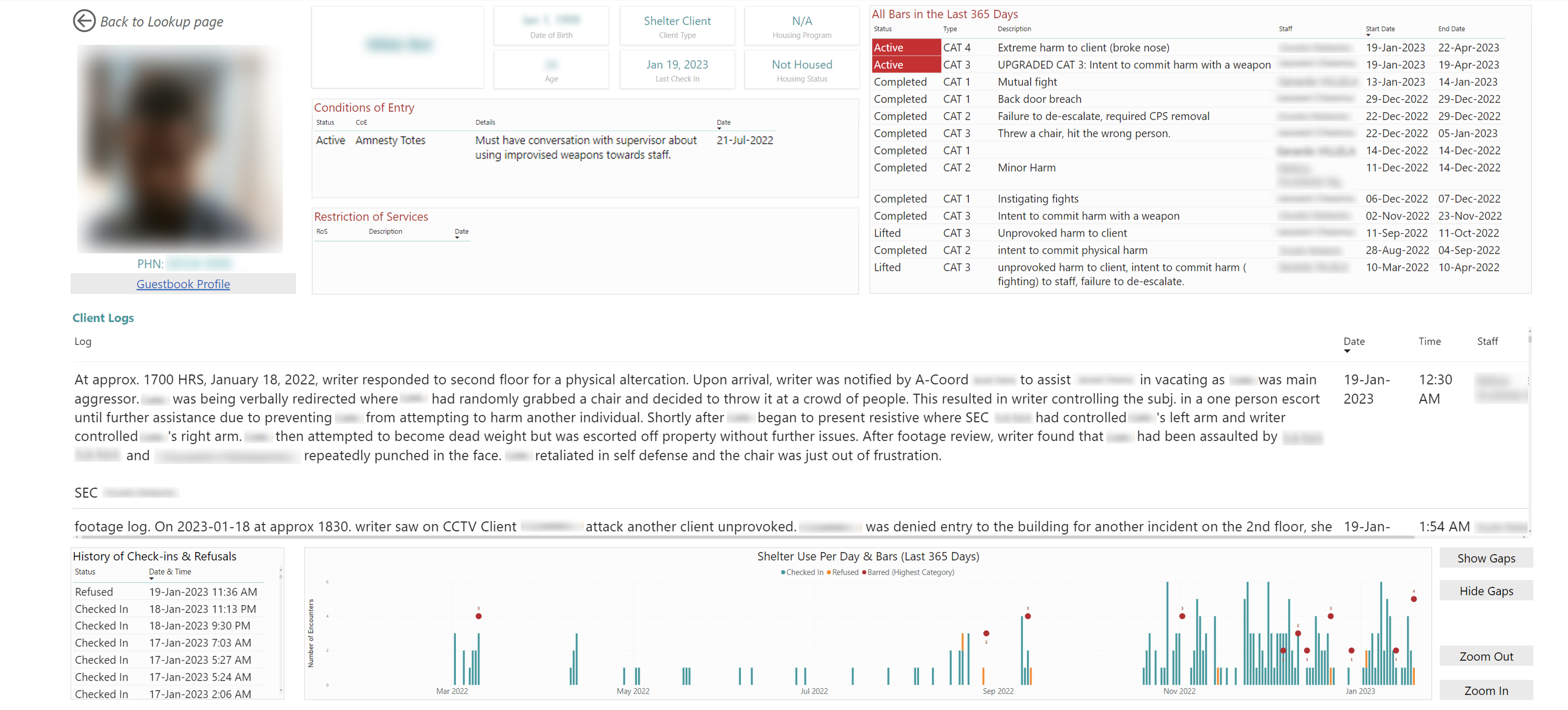}
    \caption{BRDI Screenshot: The Deep Dive Page for a selected client.}
    \Description{A screenshot of the BRDI's deep dive page for one particular client, similar to the mockup in Figure 5. The screenshot includes the same panels as in the mockup, but now with real client data with identifying information blurred for anonymity.}
    \label{fig:SSDeepDive}
\end{figure}

\revision{This appendix contains annotated mockups of the Lookup Page (Figures \ref{fig:LookupPageUnfilt} and \ref{fig:LookupPageFilt}), and the Deep Dive Page (Figure \ref{fig:DeepDive}) of BRDI V3, along with anonymous screenshots (Figures \ref{fig:SSLookup} and \ref{fig:SSDeepDive}).}

\received{October 2024}
\received[revised]{April 2025}
\received[accepted]{August 2025}

\end{document}